\documentclass[times,twocolumn]{aastex62}

\usepackage{CJK}
\usepackage{amsmath}
\usepackage{multirow}
\usepackage{cases}
\usepackage{graphicx}
\usepackage{subfigure}
\usepackage{natbib}
\usepackage{color}
\usepackage{tabularx}
\usepackage{bm}
\usepackage{threeparttable}
\usepackage{gensymb}
\usepackage{booktabs}
\usepackage{array}
\usepackage{placeins}

\newcommand{\thetae}{\theta_{\rm E}}

\newcommand{\pie}{\pi_{\rm E}}

\newcommand{\te}{t_{\rm E}}

\newcommand{\eventa}{KMT-2023-BLG-1592}
\newcommand{\eventb}{OGLE-2023-BLG-0766}
\newcommand{\eventc}{KMT-2023-BLG-0332}
\newcommand{\eventd}{KMT-2023-BLG-0486}
\newcommand{\evente}{KMT-2023-BLG-0792}
\newcommand{\eventf}{OGLE-2023-BLG-1043}

\DeclareGraphicsExtensions{.pdf,.png,.jpg}

\shorttitle{}
\shortauthors{Li et al.}

\begin{document}
\begin{CJK*}{UTF8}{gbsn}
\title{{\large Mass Production of 2023 KMTNet Microlensing Planets. II: Two Planets and A Brown Dwarf}}

\correspondingauthor{Weicheng Zang}
\email{zangweicheng@westlake.edu.cn}

\author[0000-0002-1287-6064]{Zhixing Li}
\affiliation{Department of Astronomy, Westlake University, Hangzhou 310030, Zhejiang Province, China}

\author[0009-0001-6584-7187]{Hongyu Li}
\affiliation{Department of Astronomy, Tsinghua University, Beijing 100084, China}
\affiliation{Department of Astronomy, Westlake University, Hangzhou 310030, Zhejiang Province, China}

\author[0000-0001-6000-3463]{Weicheng Zang}
\affiliation{Department of Astronomy, Westlake University, Hangzhou 310030, Zhejiang Province, China}

\author[0000-0001-9823-2907]{Yoon-Hyun Ryu} 
\affiliation{Korea Astronomy and Space Science Institute, Daejeon 34055, Republic of Korea}

\author[0000-0001-5207-5619]{Andrzej Udalski}
\affiliation{Astronomical Observatory, University of Warsaw, Al. Ujazdowskie 4, 00-478 Warszawa, Poland}

\author{Takahiro Sumi}
\affiliation{Department of Earth and Space Science, Graduate School of Science, Osaka University, Toyonaka, Osaka 560-0043, Japan}

\author[0000-0003-0626-8465]{Hongjing Yang}
\affiliation{Department of Astronomy, Westlake University, Hangzhou 310030, Zhejiang Province, China}

\author[0000-0001-5651-9440]{Yuchen Tang}
\affiliation{Department of Astronomy, Westlake University, Hangzhou 310030, Zhejiang Province, China}

\author[0000-0002-1279-0666]{Jiyuan Zhang}
\affiliation{Department of Astronomy, Tsinghua University, Beijing 100084, China}

\author[0000-0001-8317-2788]{Shude Mao}
\affiliation{Department of Astronomy, Westlake University, Hangzhou 310030, Zhejiang Province, China}

\collaboration{(Leading Authors)}

\author[0000-0003-3316-4012]{Michael D. Albrow}
\affiliation{University of Canterbury, School of Physical and Chemical Sciences, Private Bag 4800, Christchurch 8020, New Zealand}

\author[0000-0001-6285-4528]{Sun-Ju Chung}
\affiliation{Korea Astronomy and Space Science Institute, Daejeon 34055, Republic of Korea}

\author{Andrew Gould} 
\affiliation{Max-Planck-Institute for Astronomy, K\"onigstuhl 17, 69117 Heidelberg, Germany}
\affiliation{Department of Astronomy, Ohio State University, 140 W. 18th Ave., Columbus, OH 43210, USA}

\author{Cheongho Han}
\affiliation{Department of Physics, Chungbuk National University, Cheongju 28644, Republic of Korea}

\author[0000-0002-9241-4117]{Kyu-Ha Hwang}
\affiliation{Korea Astronomy and Space Science Institute, Daejeon 34055, Republic of Korea}

\author[0000-0002-0314-6000]{Youn Kil Jung}
\affiliation{Korea Astronomy and Space Science Institute, Daejeon 34055, Republic of Korea}
\affiliation{National University of Science and Technology (UST), Daejeon 34113, Republic of Korea}

\author[0000-0002-4355-9838]{In-Gu Shin}
\affiliation{Department of Astronomy, Westlake University, Hangzhou 310030, Zhejiang Province, China}

\author[0000-0003-1525-5041]{Yossi Shvartzvald}
\affiliation{Department of Particle Physics and Astrophysics, Weizmann Institute of Science, Rehovot 7610001, Israel}

\author[0000-0001-9481-7123]{Jennifer C. Yee}
\affiliation{Center for Astrophysics $|$ Harvard \& Smithsonian, 60 Garden St.,Cambridge, MA 02138, USA}

\author[0000-0002-7511-2950]{Sang-Mok Cha} 
\affiliation{Korea Astronomy and Space Science Institute, Daejeon 34055, Republic of Korea}
\affiliation{School of Space Research, Kyung Hee University, Yongin, Kyeonggi 17104, Republic of Korea} 

\author{Dong-Jin Kim}
\affiliation{Korea Astronomy and Space Science Institute, Daejeon 34055, Republic of Korea}

\author[0000-0003-0562-5643]{Seung-Lee Kim} 
\affiliation{Korea Astronomy and Space Science Institute, Daejeon 34055, Republic of Korea}

\author[0000-0003-0043-3925]{Chung-Uk Lee}
\affiliation{Korea Astronomy and Space Science Institute, Daejeon 34055, Republic of Korea}

\author[0009-0000-5737-0908]{Dong-Joo Lee} 
\affiliation{Korea Astronomy and Space Science Institute, Daejeon 34055, Republic of Korea}

\author[0000-0001-7594-8072]{Yongseok Lee} 
\affiliation{Korea Astronomy and Space Science Institute, Daejeon 34055, Republic of Korea}
\affiliation{School of Space Research, Kyung Hee University, Yongin, Kyeonggi 17104, Republic of Korea}

\author[0000-0002-6982-7722]{Byeong-Gon Park}
\affiliation{Korea Astronomy and Space Science Institute, Daejeon 34055, Republic of Korea}

\author[0000-0003-1435-3053]{Richard W. Pogge} 
\affiliation{Department of Astronomy, Ohio State University, 140 West 18th Ave., Columbus, OH  43210, USA}
\affiliation{Center for Cosmology and AstroParticle Physics, Ohio State University, 191 West Woodruff Ave., Columbus, OH 43210, USA}

\collaboration{(The KMTNet Collaboration)}

\author[0000-0001-7016-1692]{Przemek Mr\'{o}z}
\affiliation{Astronomical Observatory, University of Warsaw, Al. Ujazdowskie 4, 00-478 Warszawa, Poland}

\author[0000-0002-0548-8995]{Micha{\l}~K. Szyma\'{n}ski}
\affiliation{Astronomical Observatory, University of Warsaw, Al. Ujazdowskie 4, 00-478 Warszawa, Poland}

\author[0000-0002-2335-1730]{Jan Skowron}
\affiliation{Astronomical Observatory, University of Warsaw, Al. Ujazdowskie 4, 00-478 Warszawa, Poland}

\author[0000-0002-9245-6368]{Radoslaw Poleski}
\affiliation{Astronomical Observatory, University of Warsaw, Al. Ujazdowskie 4, 00-478 Warszawa, Poland}

\author[0000-0002-7777-0842]{Igor Soszy\'{n}ski}
\affiliation{Astronomical Observatory, University of Warsaw, Al. Ujazdowskie 4, 00-478 Warszawa, Poland}

\author[0000-0002-2339-5899]{Pawe{\l} Pietrukowicz}
\affiliation{Astronomical Observatory, University of Warsaw, Al. Ujazdowskie 4, 00-478 Warszawa, Poland}

\author[0000-0003-4084-880X]{Szymon Koz{\l}owski}
\affiliation{Astronomical Observatory, University of Warsaw, Al. Ujazdowskie 4, 00-478 Warszawa, Poland}

\author[0000-0002-9326-9329]{Krzysztof A. Rybicki}
\affiliation{Astronomical Observatory, University of Warsaw, Al. Ujazdowskie 4, 00-478 Warszawa, Poland}
\affiliation{Department of Particle Physics and Astrophysics, Weizmann Institute of Science, Rehovot 76100, Israel}

\author[0000-0002-6212-7221]{Patryk Iwanek}
\affiliation{Astronomical Observatory, University of Warsaw, Al. Ujazdowskie 4, 00-478 Warszawa, Poland}

\author[0000-0001-6364-408X]{Krzysztof Ulaczyk}
\affiliation{Department of Physics, University of Warwick, Gibbet Hill Road, Coventry, CV4~7AL,~UK}

\author[0000-0002-3051-274X]{Marcin Wrona}
\affiliation{Astronomical Observatory, University of Warsaw, Al. Ujazdowskie 4, 00-478 Warszawa, Poland}
\affiliation{Villanova University, Department of Astrophysics and Planetary Sciences, 800 Lancaster Ave., Villanova, PA 19085, USA}

\author[0000-0002-1650-1518]{Mariusz Gromadzki}
\affiliation{Astronomical Observatory, University of Warsaw, Al. Ujazdowskie 4, 00-478 Warszawa, Poland}

\author{Mateusz J. Mr\'{o}z}
\affiliation{Astronomical Observatory, University of Warsaw, Al. Ujazdowskie 4, 00-478 Warszawa, Poland}

\collaboration{(The OGLE Collaboration)}

\author{Fumio Abe}
\affiliation{Institute for Space-Earth Environmental Research, Nagoya University, Nagoya 464-8601, Japan}

\author{Ken Bando}
\affiliation{Department of Earth and Space Science, Graduate School of Science, Osaka University, Toyonaka, Osaka 560-0043, Japan}

\author{David P. Bennett}
\affiliation{Code 667, NASA Goddard Space Flight Center, Greenbelt, MD 20771, USA}
\affiliation{Department of Astronomy, University of Maryland, College Park, MD 20742, USA}

\author{Aparna Bhattacharya}
\affiliation{Code 667, NASA Goddard Space Flight Center, Greenbelt, MD 20771, USA}
\affiliation{Department of Astronomy, University of Maryland, College Park, MD 20742, USA}

\author{Ian A. Bond}
\affiliation{Institute of Natural and Mathematical Sciences, Massey University, Auckland 0745, New Zealand}

\author{Akihiko Fukui}
\affiliation{Department of Earth and Planetary Science, Graduate School of Science, The University of Tokyo, 7-3-1 Hongo, Bunkyo-ku, Tokyo 113-0033, Japan}
\affiliation{Instituto de Astrof\'isica de Canarias, V\'ia L\'actea s/n, E-38205 La Laguna, Tenerife, Spain}

\author{Ryusei Hamada}
\affiliation{Department of Earth and Space Science, Graduate School of Science, Osaka University, Toyonaka, Osaka 560-0043, Japan}

\author{Shunya Hamada}
\affiliation{Department of Earth and Space Science, Graduate School of Science, Osaka University, Toyonaka, Osaka 560-0043, Japan}

\author{Naoto Hamasak}
\affiliation{Department of Earth and Space Science, Graduate School of Science, Osaka University, Toyonaka, Osaka 560-0043, Japan}

\author{Yuki Hirao}
\affiliation{Department of Earth and Space Science, Graduate School of Science, Osaka University, Toyonaka, Osaka 560-0043, Japan}

\author{Stela Ishitani Silva}
\affiliation{Department of Physics, The Catholic University of America, Washington, DC 20064, USA}
\affiliation{Code 667, NASA Goddard Space Flight Center, Greenbelt, MD 20771, USA}

\author{Naoki Koshimoto}
\affiliation{Department of Astronomy, University of Maryland, College Park, MD 20742, USA}

\author{Yutaka Matsubara}
\affiliation{Institute for Space-Earth Environmental Research, Nagoya University, Nagoya 464-8601, Japan}

\author{Shota Miyazaki}
\affiliation{Department of Earth and Space Science, Graduate School of Science, Osaka University, Toyonaka, Osaka 560-0043, Japan}

\author{Yasushi Muraki}
\affiliation{Institute for Space-Earth Environmental Research, Nagoya University, Nagoya 464-8601, Japan}

\author{Tutumi Nagai}
\affiliation{Institute for Space-Earth Environmental Research, Nagoya University, Nagoya 464-8601, Japan}

\author{Kansuke Nunota}
\affiliation{Institute for Space-Earth Environmental Research, Nagoya University, Nagoya 464-8601, Japan}

\author{Greg Olmschenk}
\affiliation{Code 667, NASA Goddard Space Flight Center, Greenbelt, MD 20771, USA}

\author{Cl\'ement Ranc}
\affiliation{Sorbonne Universit\'e, CNRS, Institut d'Astrophysique de Paris, IAP, F-75014, Paris, France}

\author{Nicholas J. Rattenbury}
\affiliation{Department of Physics, University of Auckland, Private Bag 92019, Auckland, New Zealand}

\author{Yuki Satoh}
\affiliation{Department of Earth and Space Science, Graduate School of Science, Osaka University, Toyonaka, Osaka 560-0043, Japan}

\author{Daisuke Suzuki}
\affiliation{Department of Earth and Space Science, Graduate School of Science, Osaka University, Toyonaka, Osaka 560-0043, Japan}

\author{Sean Terry}
\affiliation{Code 667, NASA Goddard Space Flight Center, Greenbelt, MD 20771, USA}
\affiliation{Department of Astronomy, University of Maryland, College Park, MD 20742, USA}

\author{Paul J. Tristram}
\affiliation{University of Canterbury Mt.\ John Observatory, P.O. Box 56, Lake Tekapo 8770, New Zealand}

\author{Aikaterini Vandorou}
\affiliation{Code 667, NASA Goddard Space Flight Center, Greenbelt, MD 20771, USA}
\affiliation{Department of Astronomy, University of Maryland, College Park, MD 20742, USA}

\author{Hibiki Yama}
\affiliation{Department of Earth and Space Science, Graduate School of Science, Osaka University, Toyonaka, Osaka 560-0043, Japan}

\collaboration{(The MOA Collaboration)}

\begin{abstract}

To expand the homogeneous microlensing planetary sample of the Korea Microlensing Telescope Network (KMTNet), we investigate six planetary candidates identified by the AnomalyFinder search in the 2023 prime-field data, namely KMT-2023-BLG-1592, OGLE-2023-BLG-0766, KMT-2023-BLG-0332, KMT-2023-BLG-0486, KMT-2023-BLG-0792, and OGLE-2023-BLG-1043. Light-curve modeling indicates that the first two events have planetary mass ratios of $\log q \sim -3.0$ and $-2.6$, while the third exhibits a brown dwarf mass ratio of $\log q \sim -1.4$. The remaining three events show the well-known degeneracy between the binary-lens single-source (2L1S) and single-lens binary-source (1L2S) interpretations. A Bayesian analysis yields companion masses of about 0.6 and 1.2 Jupiter masses for the two planetary systems, likely orbiting beyond the snow lines of M- or K-dwarf hosts. A review of the KMTNet planetary sample shows that candidates discovered by AnomalyFinder are significantly more likely to exhibit the 2L1S/1L2S degeneracy, consistent with the tendency of AnomalyFinder to detect subtler planetary signals.

\end{abstract}

\section{Introduction}\label{sec:intro}

Microlensing offers a uniquely powerful and complementary pathway to exoplanet discovery \citep{Shude1991,Andy1992}. In contrast to the transit and radial-velocity methods, which account for the vast majority of known planets, microlensing is intrinsically sensitive to cold planets at or beyond Jupiter-like separations across the full range of planetary masses, and even to unbound planetary-mass objects \citep{Sumi2011,Mroz2017a,Gould2022_FFP_EinsteinDesert,Sumi2023}. Because the microlensing signal is independent of light from the host star, it enables the detection of planets orbiting faint or dark hosts, including brown dwarfs (e.g., \citealt{OB120358}) and white dwarfs \citep{MB10477_AO,KB200414_AO}, as well as planets located in distant stellar populations throughout the Milky Way (e.g., \citealt{Matthewbulge,OB161190}).

To date, six homogeneously selected statistical samples have been constructed for microlensing searches for bound planets \citep{mufun,Cassan2012,Suzuki2016,Wise,OGLE_wide,OB160007}. Among them, \cite{OB160007} presents the largest sample, comprising 63 planets discovered by the Korea Microlensing Telescope Network (KMTNet; \citealt{KMT2016}) and its AnomalyFinder system \citep{OB191053,2019_prime}. Their analysis reveals two distinct populations of microlensing planets, a population of gas giants and a population of super-Earths/mini-Neptunes, and identifies a possible deficit in planet-to-host mass ratio ($q$) at $-3.6 < \log q < -3.0$. This potential ``mass-ratio desert'' may reflect the result of runaway gas accretion during planet formation \citep{Ida2004,Mordasini2009}. The feature was later strengthened by investigating all KMTNet AnomalyFinder planets from 2016 to 2019 \citep{2017_subprime}, although it was not apparent in the Microlensing Observations in Astrophysics (MOA, \citealt{Sako2008}) statistical sample, which contains 22 planets \citep{Suzuki2016}.

To further test the results of \cite{OB160007}, a larger sample is required. KMTNet monitors approximately $97~\mathrm{deg}^2$ of the Galactic bulge, including $\sim 13~\mathrm{deg}^2$ of prime fields with cadences of $\Gamma \geq 2~\mathrm{hr}^{-1}$ and $\sim 84~\mathrm{deg}^2$ of subprime fields with cadences of $\Gamma \leq 1~\mathrm{hr}^{-1}$. The field layout and cadence distribution are shown in Figure~12 of \cite{KMTeventfinder}. KMTNet identifies microlensing events through two channels. During the observing season, the KMTNet AlertFinder system \citep{KMTAF} scans the data every weekday to detect ongoing events in real time. In addition, after each season, the EventFinder system \cite{KMTeventfinder} analyzes the full yearly dataset and recovers hundreds of events missed by the real-time alerts. All discovered events are processed by the automatic pySIS pipeline \citep{pysis}, yielding light curves based on the  difference imaging analysis (DIA) technique \citep{Tomaney1996,Alard1998}.

Shin et al.\ (2026) have presented the complete KMTNet sample from all 2021 prime-field events, while Shin et al.\ (in prep.) and Ryu et al.\ (in prep.) will report results from the 2021 subprime- and 2022 prime-field datasets, respectively. For the 2023 season, we re-reduced all events with the optimized pySIS pipeline \citep{Yang_TLC,Yang_TLC2} and subsequently ran the AnomalyFinder on these reprocessed data. Besides published planets from by-eye searches (e.g., \citealt{KB231866,KB230416,KB230119}), Ryu et al.\ (2026) have presented the full sample of low mass-ratio planets with $\log q < -3.7$ from the 2023 AlertFinder events. In this paper we present detailed analysis of six events, which reveals two secure planets, one brown dwarf, and three ambiguous candidates drawn from the 2023 prime-field AlertFinder sample. In addition, Li et al.\ (in prep.) will introduce an additional six events from the 2023 subprime-field AlertFinder sample. The planetary sample derived from the 2023 EventFinder search will be published in forthcoming papers.

\section{Observations}\label{sec:obser}

\begin{table*}
    \renewcommand\arraystretch{1.35}
    \centering
    \caption{Event Names, Alert, Locations, and Cadences for the five events analyzed in this paper}
    \begin{tabular}{c c c c c c c c}
    \hline
    \hline
    Event Name & First Alert Date & ${\rm RA}_{\rm J2000}$ & ${\rm Decl.}_{\rm J2000}$ & $\ell$ & $b$ & Cadence & Method \\
    \hline
    \eventa & 10 Jul 2023 & 17:54:42.88 & $-$29:00:30.31 & +0.9583 & $-$1.7446 & $2~{\rm hr}^{-1}$ & Recovery \\
    \hline
    \eventb & 17 Jun 2023 & 17:59:19.11 & $-$28:19:28.88 & +2.0565 & $-$2.2763 & $3~{\rm night}^{-1}$ & Discovery \\
    KMT-2023-BLG-1287 & & & & & & $4~{\rm hr}^{-1}$ \\
    \hline 
    \eventc & 10 Apr 2023 & 18:03:33.63 & $-$27:13:20.60 & +3.4796 & $-$2.5481 & $4~{\rm hr}^{-1}$ & Discovery \\
    OGLE-2023-BLG-0317 & & & & & & $3~{\rm night}^{-1}$ \\
    \hline
    \eventd & 20 Apr 2023 & 17:58:40.28& $-$26:54:34.60 & +3.2130 & $-$1.4474	 & $2~{\rm hr}^{-1}$ & Discovery \\
    \hline
    \evente & 12 May 2023 & 17:56:44.25 & $-$30:36:55.12 & $-$0.2114 & $-$2.9315 & $4~{\rm hr}^{-1}$ & Discovery \\
    OGLE-2023-BLG-0604 & & & & & & $1~{\rm night}^{-1}$ \\
    MOA-2023-BLG-244 & & & & & & $4~{\rm hr}^{-1}$ \\
    \hline
    \eventf & 04 Aug 2023 & 17:53:38.45 & $-$28:53:51.61 & +0.9350 & $-$1.4860  & $1~{\rm hr}^{-1}$ & Discovery \\
    KMT-2023-BLG-1951 & & & & & & $4~{\rm hr}^{-1}$\\
    MOA-2023-BLG-395 & & & & & & $4~{\rm hr}^{-1}$ \\
    \hline
    \hline
    \end{tabular}
     \tablecomments{``Discovery'' indicates that the planet was discovered using AnomlyFinder, and ``Recovery'' means that the planet was first discovered from by-eye searches and then recovered by AnomlyFinder.}
    \label{event_info}
\end{table*}

Of the six events, four events, \eventa, KMT-2023-BLG-0332/OGLE-2023-BLG-0317, \eventc, and KMT-2023-BLG-0792/OGLE-2023-BLG-0604/MOA-2023-BLG-244, were first flagged by the KMTNet AlertFinder system. The remaining two events, OGLE-2023-BLG-0766/KMT-2023-BLG-1287 and OGLE-2023-BLG-1043/KMT-2023-BLG-1951/MOA-2023-BLG-395, were initially discovered by the Early Warning System \citep{Udalski1994, Udalski2003} of the Optical Gravitational Lensing Experiment (OGLE; \citealt{OGLEIV}). Subsequently, the MOA survey also reported two events.

As described earlier, the KMTNet survey obtained data using its three telescopes located at KMTC, KMTS, and KMTA. The OGLE collaboration carried out observations from its 1.3\,m telescope in Chile, which is equipped with a wide-field camera covering $1.4~{\rm deg}^2$ \citep{OGLEIV}. The MOA survey conducted observations with a 1.8\,m telescope in New Zealand, featuring a camera with a field of view of $2.2~{\rm deg}^2$.

The majority of the KMTNet and OGLE data were acquired in the $I$ band, whereas MOA observations were mainly carried out using the MOA-Red filter, a broad passband approximately equivalent to the combined Cousins $R$ and $I$ bands. In addition, each survey obtained a smaller number of images in the $V$ band, which were used to determine the color of the source stars.

Table~\ref{event_info} provides an overview of the six microlensing events analyzed in this work, listing their event names, initial alert dates, equatorial and Galactic coordinates, the observing cadences of the participating surveys, as well as the anomaly identification method. Throughout this paper, each event is referred to by the designation under which it was first reported.

For the light-curve modeling, we adopted the photometric data products produced by the difference image analysis (DIA) pipelines of each survey: \cite{pysis, Yang_TLC, Yang_TLC2} for KMTNet, \cite{Wozniak2000} for OGLE, and \cite{Bond2001} for MOA. The photometric error bars were subsequently rescaled following the method described in \cite{MB11293}, such that each data set yields a reduced $\chi^{2}$ of unity.

\section{Light-curve Analysis}\label{sec:model}

\begin{table*}[htb]
  \renewcommand\arraystretch{1.20}
  \centering
  \caption{Lensing Parameters for \eventa}
  \begin{tabular}{c|c c c}
    \hline\hline
    Parameter & Wide & Close & 1L2S \\
    \hline
    $\chi^2$/dof &6756.0/6756  &6752.2/6756  &7607.6/6756\\ \hline
    $t_0$ (HJD$'$) &$10134.259 \pm 0.001$  &$10134.260\pm0.001$  &$10134.217\pm0.001$\\
    $u_0 (10^{-4})$ &$5.38 \pm 1.37$  &$5.17 \pm 1.19$  &$5.48\pm0.84$\\
    $t_{\rm E}$ (days) & $80.4\pm17.3$ &$91.7\pm21.0$  &$70.6\pm11.6$\\
    $\rho (10^{-4})$ &$7.00\pm 1.54$  &$6.15\pm 1.41$  & ---\\
    $\alpha$ (deg) &$122.54 \pm 0.47$  &$122.19 \pm 0.49$  & --- \\
    $s$ &$1.399\pm0.007$  &$0.700\pm0.005$ & --- \\
    $\log q$ &$-3.010\pm0.096$ & $-3.017\pm0.098$  & --- \\
    $t_{0,2}$ (HJD$'$) & --- & --- & $10134.508\pm  0.004$ \\
    $u_{0,2} (10^{-4})$ & --- & --- & $11.39 \pm 1.93$ \\
    $q_{f, I}$ & --- & --- & $ 0.3739\pm 0.0215$ \\
    $I_{\rm S,KMTC}$ &$23.389\pm0.268$  &$23.406\pm0.242$  &$23.284\pm0.188$\\
    \hline\hline
  \end{tabular}
  \tablecomments{HJD$^\prime = \mathrm{HJD} -2450000$.}
  \label{tab:parm-a}
\end{table*}

\begin{figure}
    \centering
    \includegraphics[width=0.47\textwidth]{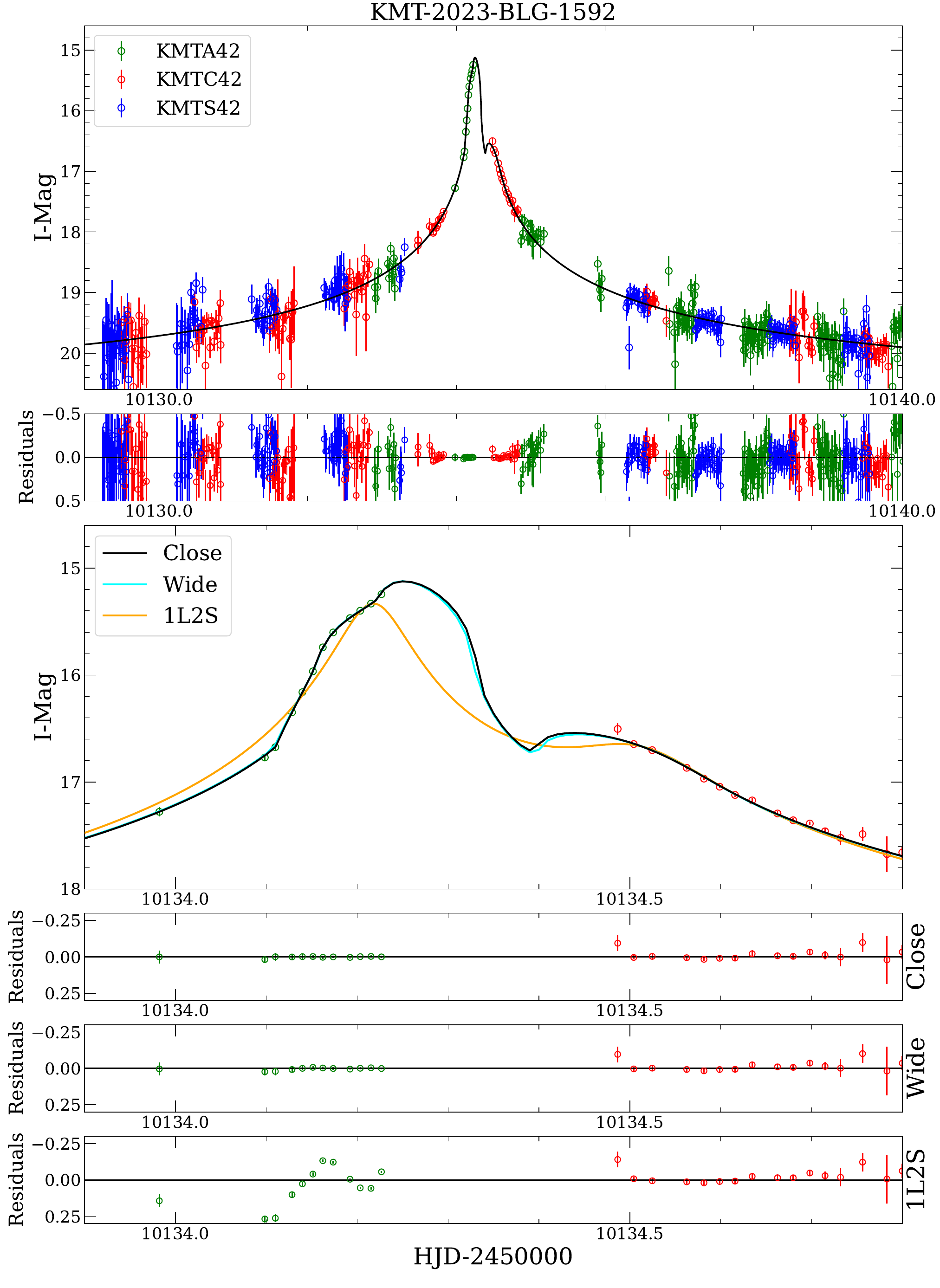}
    \caption{Light curves of \eventa\ with the 2L1S ``Close'', ``Wide'' and the 1L2S models shown as solid black, cyan and yellow curves, respectively. The lower panel provides a zoom-in view of the anomaly. Data from different observatories are plotted in different colors. The corresponding lensing model parameters are listed in Table~\ref{tab:parm-a}.}
\label{fig:lc-a}
\end{figure}

\begin{figure*}
    \centering
    \includegraphics[width=0.97\textwidth]{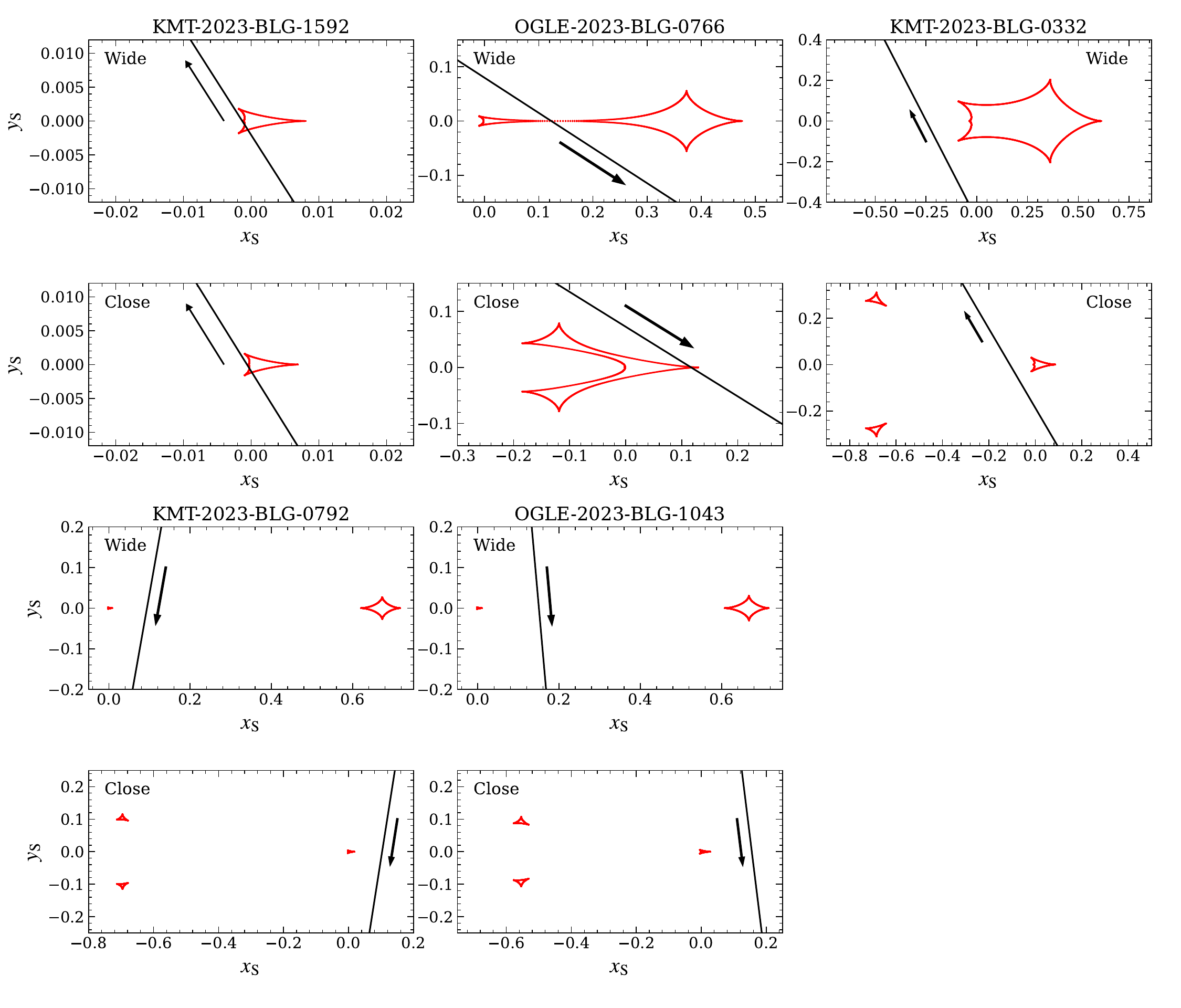}
    \caption{Geometries of the 2L1S solutions for \eventa, \eventb, \eventc, \evente, and \eventf, each of which exhibits a pair of ``Close/Wide'' degeneracies. For each event, the upper and lower panels display the corresponding ``Wide'' and ``Close'' solutions. In every panel, the red curves indicate the caustic structures, the black curve traces the source trajectory, and the arrow marks the direction of source motion.}
\label{fig:cau_abcef}
\end{figure*}

\subsection{Preamble}\label{model_preamble}

In this section, we present the methodology used to model the observed microlensing light curves. While our overall strategy follows the general framework adopted in previous analyses of anomalous microlensing events (e.g., \citealt{-4planet}), the implementation details and emphasis differ in several aspects. To maintain clarity and avoid unnecessary repetition, we provide here a concise but self-contained description of the modeling procedure, and refer interested readers to \cite{-4planet} for a more comprehensive background discussion.

All events are firstly analyzed under the assumption of a static binary-lens single-source (2L1S) configuration. The model is specified by seven parameters. Three of these correspond to the standard point-source point-lens (PSPL, \citealt{Paczynski1986}) description: the epoch of closest lens-source approach $t_0$, the minimum impact parameter $u_0$ expressed in units of the angular Einstein radius $\theta_{\rm E}$, and the Einstein timescale $\te$. The latter is related to the physical properties of the lens system through
\begin{equation}
\te = \frac{\theta_{\rm E}}{\mu_{\rm rel}}, \qquad
\theta_{\rm E} = \sqrt{\kappa M_{\rm L} \pi_{\rm rel}},
\end{equation}
where $\kappa = 4G/(c^2\,{\rm au}) \simeq 8.144~{\rm mas}\,M_{\odot}^{-1}$, $M_{\rm L}$ denotes the total lens mass, and $\pi_{\rm rel}$ and $\mu_{\rm rel}$ are the lens-source relative parallax and proper motion, respectively.

Three parameters characterize the binary nature of the lens. These include the projected separation $s$ between the two lens components normalized to $\theta_{\rm E}$, the mass ratio $q$, and the angle $\alpha$ between the source trajectory and the binary axis. Finally, finite-source effects \citep{1994ApJ...421L..75G,Shude1994,Nemiroff1994} are incorporated through the normalized source radius $\rho = \theta_*/\theta_{\rm E}$, where $\theta_*$ is the angular radius of the source star.

Model light curves are computed using the contour-integration algorithm implemented in the \texttt{VBBinaryLensing} package \citep{Bozza2010,Bozza2018,VBMicrolensing2025}. For each observatory data set $i$, the observed flux $f_i(t)$ is modeled as a linear combination of the magnified source flux and an additional blended component,
\begin{equation}
f_i(t) = f_{{\rm S},i}\,A(t) + f_{{\rm B},i},
\end{equation}
where $f_{{\rm S},i}$ and $f_{{\rm B},i}$ represent the source and blend fluxes, respectively.

To identify viable solutions, we combine a coarse and a dense exploration of parameter space using a multidimensional grid in $(\log q, \log s, \rho)$, while allowing $(t_0, u_0, \te, \alpha)$ to vary continuously. Local minima are then refined through a Markov Chain Monte Carlo (MCMC) analysis of the $\chi^2$ surface using the affine-invariant ensemble sampler \texttt{emcee} \citep{emcee2,emcee}. The final parameter optimization is performed via downhill minimization with \texttt{SciPy} \citep{scipy}. In the vicinity of each local minimum, all parameters are allowed to vary freely. Parameter estimates reported throughout this work correspond to posterior median values, with uncertainties given by the central 68\% credible intervals.

We further investigate whether annual microlensing parallax effects provide meaningful constraints for any of the events \citep{Gould1992,Gould2000,Gouldpies2004}. The microlensing parallax vector is defined as
\begin{equation}
\boldsymbol{\pi}_{\rm E} = \frac{\pi_{\rm rel}}{\theta_{\rm E}} \frac{\boldsymbol{\mu}_{\rm rel}}{\mu_{\rm rel}},
\end{equation}
and is parameterized by its north and east components $(\pi_{\rm E,N}, \pi_{\rm E,E})$. When parallax is included in the modeling, we also account for possible orbital motion of the lens system \citep{MB09387,OB09020} and explicitly explore both $u_0>0$ and $u_0<0$ solutions in order to capture the well-known ecliptic degeneracy \citep{Jiang2004,Poindexter2005}.

Because all events are characterized by smooth, bump-like deviations rather than dip-like anomalies, we additionally test a single-lens binary-source (1L2S) interpretation \citep{Gaudi1998}. In this scenario, the observed magnification in a given photometric band $\lambda$ is the flux-weighted sum of the magnifications of the two source stars \citep{MB12486},
\begin{equation}
A_{\lambda}(t) = \frac{A_{1}(t) f_{{\rm S},1,\lambda} + A_{2}(t) f_{{\rm S},2,\lambda}}
{f_{{\rm S},1,\lambda}+f_{{\rm S},2,\lambda}}
= \frac{A_{1}(t) + q_{f,\lambda} A_{2}(t)}{1 + q_{f,\lambda}},
\end{equation}
where $q_{f,\lambda} = f_{{\rm S},2,\lambda}/f_{{\rm S},1,\lambda}$ is the source flux ratio and $A_j(t)$ ($j=1,2$) denotes the magnification of each source.

Following \cite{OB160007}, several studies have adopted $\Delta\chi^2<10$ as a criterion for identifying degenerate solutions (see also, e.g., \citealt{2016_prime}). In this work, we apply a more conservative threshold and exclude models with $\Delta\chi^2>20$. Because the $\Delta\chi^2$ values for all candidate solutions are explicitly reported, readers may readily impose alternative criteria depending on the requirements of specific statistical analyses or population studies.

Below, we first present three unambiguous 2L1S events and then introduce three additional events that can be described by both 2L1S and 1L2S models.

\subsection{\eventa}

The anomaly of \eventa, as shown in Figure~\ref{fig:lc-a}, is a bump-type feature near the peak, captured by the KMTA42 data. The grid search finds only a pair of solutions, although there is a 0.25~day gap between the KMTA42 and KMTC42 data over the peak. Following standard convention, we denote these solutions as ``Close'' and ``Wide'' corresponding to projected separations of $s<1$ and $s>1$, respectively \citep{Griest1998}. The 2L1S parameters for each solution are presented in Table~\ref{tab:parm-a}, and the $\Delta\chi^2$ between the two solutions is only 3.8. As shown in the lower panels of Figure~\ref{fig:lc-a}, both solutions provide good fits to the data. The caustic topology is shown in Figure~\ref{fig:cau_abcef}. The anomaly is produced by a source caustic crossing of the central caustic.

The source star is very faint, with $I_{\rm S} = 23.4$, so the peak magnification reaches $\sim 2100$, making this one of the highest-magnification microlensing events \citep{OB04343,OB07224,OB08279,KB200414}, although the uncertainty of the source magnitude is large. Finite-source effects are detected in both solutions, although the data cover only about half of the anomaly.

Because the anomaly shows a secondary rise rather than a single bump, the 1L2S model is strongly rejected, with $\Delta\chi^2 = 855$. Due to the very faint source, this event provides no useful constraint on the parallax, with 1-$\sigma$ uncertainties in the parallax vector $0.4$ in all directions. With $\log q \sim -3.0$, this event reveals a new gas giant. 

\subsection{\eventb}

\begin{figure}
    \centering
    \includegraphics[width=0.47\textwidth]{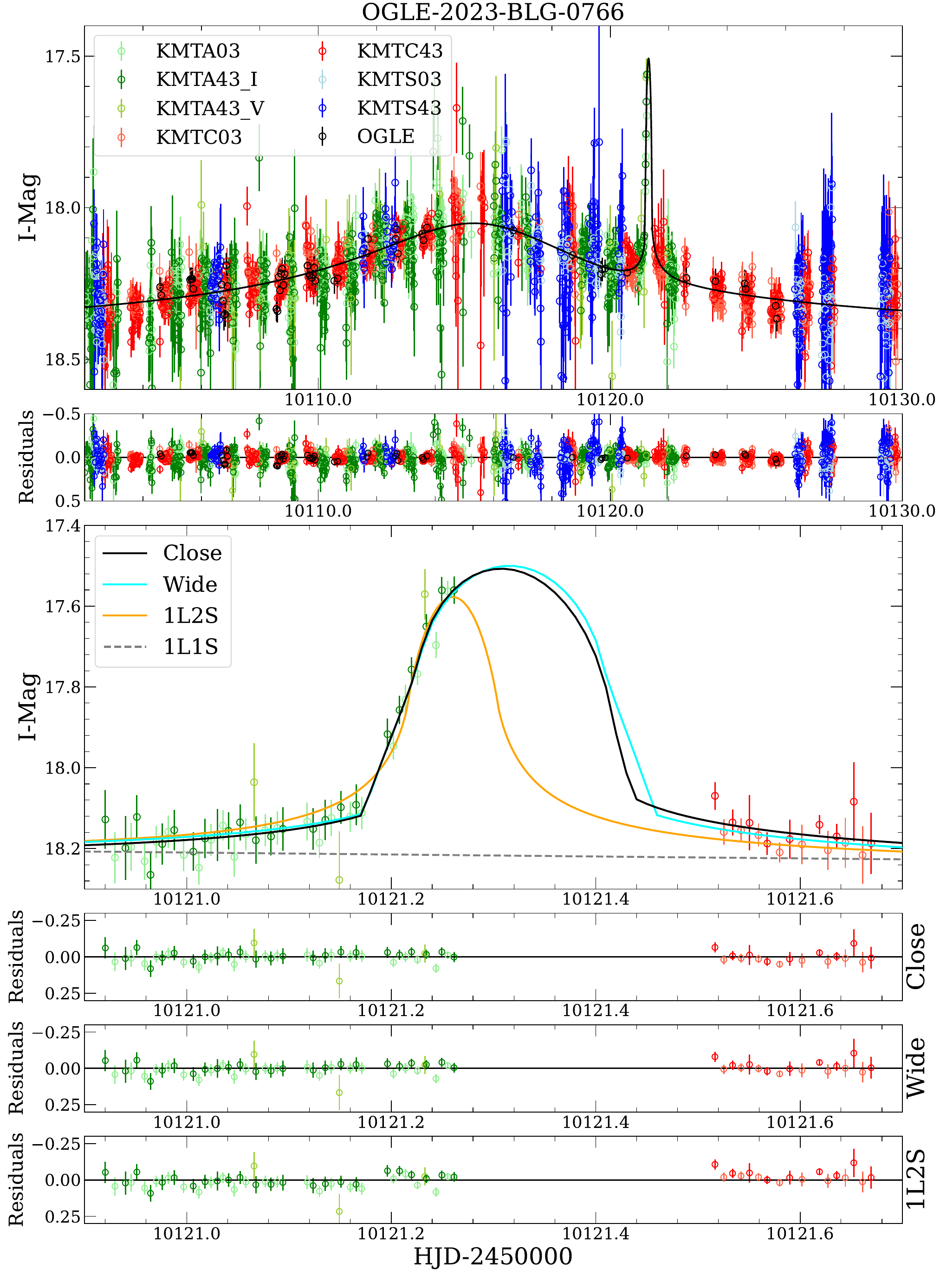}
    \caption{The observed data and models for \eventb. The symbols are similar to those in Figure \ref{fig:lc-a}.}
\label{fig:lc-b}
\end{figure}

\begin{table*}[htb]
\renewcommand\arraystretch{1.20}
\centering
\caption{Lensing Parameters for \eventb}
\begin{tabular}{c | c c c}
\hline\hline
Parameters & Wide & Close & 1L2S \\
\hline
$\chi^2$/dof & 14084.7/14088 & 14084.3/14088 & 14109.8/14088 \\ \hline
$t_0$ (HJD$'$) & $10115.56 \pm 0.06$ & $10115.57 \pm 0.05 $ & $10115.08\pm 0.05$\\
$u_0$ & $0.0731 \pm 0.0011$ & $0.0669 \pm 0.0067$ & $ 0.0519\pm 0.0080$ \\
$t_{\rm E}$ (days) & $52.2 \pm 6.1$ & $55.0\pm 3.8$ & $69.5\pm 10.2$ \\
$\rho (10^{-3})$ & $1.14 \pm 0.17$ & $1.00 \pm 0.17$ &  --- \\
$\alpha$ (deg) & $326.89 \pm 0.76$ & $327.50\pm 0.88$ & --- \\
$s$ & $1.204 \pm 0.026$ & $0.942 \pm 0.011$ & --- \\
$\log q$ & $-2.564 \pm 0.036$ & $-2.589 \pm  0.045$ & --- \\
$t_{0,2}$ (HJD$'$) & --- & --- & $10121.26\pm  0.01$ \\
$u_{0,2} (10^{-4})$ & --- & --- & $1.00 \pm 0.70$ \\
$\rho_2 (10^{-3})$ & --- & --- & $0.66\pm0.10$ \\
$q_{f, I}$ & --- & --- & $0.0131\pm 0.0012$ \\
$q_{f, V}$ & --- & --- & $0.0180\pm 0.0054$ \\
$I_{\rm S,KMTC}$ & $ 22.058\pm 0.183$ & $ 22.034\pm 0.112$ & $ 22.658\pm 0.187$ \\
\hline\hline
\end{tabular}
\label{tab:parm-b}
\end{table*}

As shown in Figure~\ref{fig:lc-b}, the anomaly of \eventb\ appears as a short-lived bump about six days after the main peak. It is primarily characterized by the KMTA data, with additional support from KMTC. The grid search yields two local minima. Their caustic geometries and source trajectories are shown in Figure~\ref{fig:cau_abcef}, and the corresponding 2L1S lensing parameters are listed in Table~\ref{tab:parm-b}. The anomaly is produced by a source crossing of a resonant caustic, and we therefore refer to the two solutions as the ``Close'' and ``Wide'' models. Both solutions provide similarly good fits to the data, with a $\chi^2$ difference of only 0.4. Finite-source effects are measured in both cases.

The 1L2S model is disfavored by $\Delta\chi^2 = 25.5$, which is statistically significant. As shown in Figure~\ref{fig:lc-b}, the 1L2S model mainly fails to reproduce the wings of the anomaly. In addition, the 1L2S model could be verified by searching for a color change during the anomaly, since a difference in the colors of the two sources would produce chromatic variations \citep{Shude1991,Gaudi1998}. A single KMTA43 $V$-band data point obtained near the peak of the anomaly constrains the color of the secondary source. As discussed in Section~\ref{sec:phy_eventb}, the inferred color places the putative secondary source away from the main-sequence branch, further excluding the 1L2S interpretation.

Due to the faintness of the event, the microlensing parallax is poorly constrained, with 1-$\sigma$ uncertainties in the parallax vector $>1$ in all directions. We therefore adopt the static 2L1S model as the final interpretation of this event. With $\log q \sim -2.6$, the lens companion is consistent with a gas-giant planet.

\subsection{\eventc}

\begin{figure}
    \centering
    \includegraphics[width=0.47\textwidth]{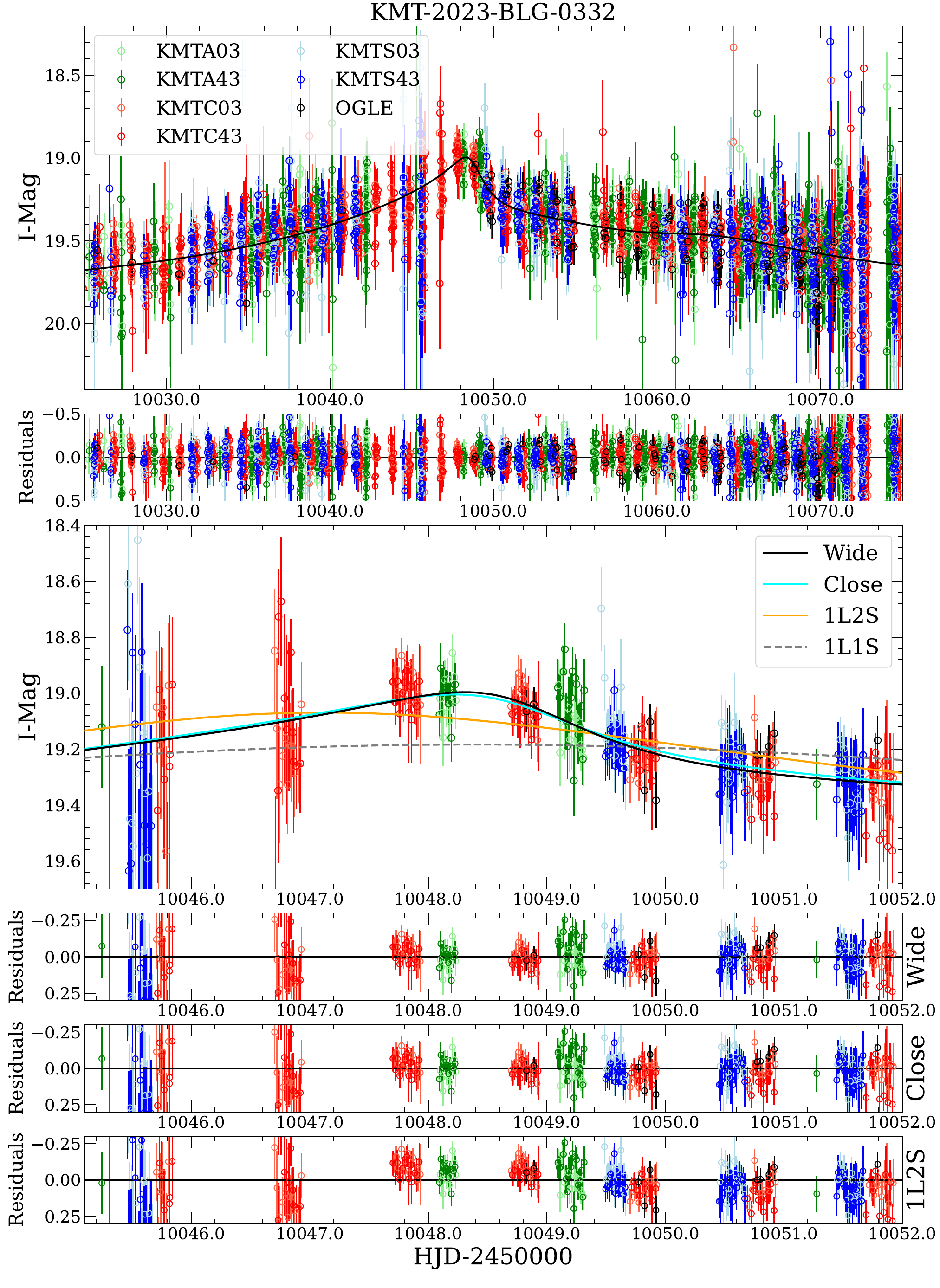}
    \caption{The observed data and models for \eventc. The symbols are similar to those in Figure \ref{fig:lc-a}.}
\label{fig:lc-c}
\end{figure}

The anomaly of \eventc, as shown in Figure~\ref{fig:lc-c}, is an asymmetric feature near the peak, captured by both the KMTNet and OGLE data. The grid search yields a pair of ``Close/Wide'' solutions. As shown in Figure~\ref{fig:cau_abcef}, the anomaly is produced by a cusp approach to the central caustic (for ``Close'') or to the resonant caustic (for ``Wide''). According to the lensing parameters listed in Table~\ref{tab:parm-c}, the ``Wide'' solution provides the best fit, with a brown dwarf mass ratio of $\log q = -1.404 \pm 0.054$. The ``Close'' solution has a planetary mass ratio of $\log q = -1.732 \pm 0.029$ and is disfavored by $\Delta\chi^2 = 30$, which is sufficient to reject it. No useful constraint on finite-source effects is obtained, with the 1-$\sigma$ upper limit on $\rho$ exceeding 0.01.

The 1L2S model, whose parameters are also listed in Table~\ref{tab:parm-c}, is strongly excluded, with $\Delta\chi^2 = 219$. Because of the extended demagnified region following the cusp approach, the anomaly does not exhibit a typical bump-type morphology. Due to the faintness of the event, with a peak brightness of only $I \sim 19$ mag, no meaningful constraint on parallax can be derived from the data.

With $\log q = -1.404 \pm 0.054$, the companion is likely a brown dwarf. Therefore, this event is not included in the study of the planetary mass-ratio function, but it can be incorporated into future statistical analysis of brown dwarf companions.

\begin{table*}[htb]
    \renewcommand\arraystretch{1.20}
    \caption{Lensing parameters for \eventc}
    \centering
        \begin{tabular}{c|ccc}
        \hline\hline
        Parameter & Wide & Close & 1L2S \\
        \hline
        $\chi^2/\mathrm{dof}$ 
        & 13106.5/13110 & 13136.5/13110 & 13325.4/13110 \\ \hline
        $t_0$ (HJD$^\prime$) 
        & $10050.50\pm0.19$ & $10049.68\pm0.12$ & $ 10047.02\pm0.09$ \\
        $u_0$ 
        & $0.2061\pm0.0309$ & $0.0908\pm0.0079$ & $ 0.0279\pm0.0066$ \\
        $t_{\rm E}$ (days) 
        & $45.23\pm4.96$ & $81.16\pm6.93$ & $146.70\pm 32.48$ \\
        $\alpha$ (deg) 
        & $117.48\pm0.80$ & $120.64\pm0.63$ & ---\\
        $s$ 
        & $1.197\pm0.032$ & $0.705\pm0.004$ & --- \\
        $\log q$ 
        & $-1.404\pm0.054$ & $-1.732\pm0.029$ & --- \\
        $t_{0,2}$ (HJD$'$) & --- & --- & $10063.49\pm 0.41$ \\
        $u_{0,2}$ & --- & --- & $0.0412 \pm 0.0119$ \\
        $q_{f, I}$ & --- & --- & $0.3375\pm 0.0509$ \\
        $I_{\rm S,KMTC}$ 
        & $21.508\pm0.186$ & $22.429\pm0.096$ & $23.669\pm0.346$ \\
        \hline\hline
        \end{tabular}
        \label{tab:parm-c}
\end{table*}

\subsection{\eventd}

\begin{table*}[htb]
  \renewcommand\arraystretch{1.20}
  \centering
  \caption{Lensing Parameters for \eventd}
  \begin{tabular}{c|c c c c}
    \hline
    \hline
    Parameter & Wide & Close & Off-axis & 1L2S\\
    \hline
    $\chi^2$/dof & 6296.4/6297 & 6296.1/6297 & 6327.2/6297 & 6290.6/6297 \\ \hline
    $t_0$ (HJD$'$) & $10078.33 \pm 0.03$ & $10078.26 \pm 0.04$ & $10078.96 \pm 0.08 $ & $10078.60\pm 0.03$\\
    $u_0$ & $0.0240 \pm 0.0063$ & $0.0396 \pm 0.0079$ & $-0.0625 \pm 0.0030$ & $ 0.0412\pm 0.0046$ \\
    $t_{\rm E}$ (days) & $217.8 \pm  61.9$ & $128.9 \pm 28.5$ & $82.8\pm 3.9$ & $118.3\pm 13.6$ \\
    $\rho (10^{-2})$ & --- & --- & $2.73 \pm 0.76$ &  --- \\
    $\alpha$ (deg) & $199.1 \pm 0.3$ & $198.9 \pm 0.4$ & $5.63\pm 0.82$ & --- \\
    $s$ & $1.224 \pm 0.018$ & $0.882 \pm 0.016$ & $0.912 \pm 0.021$ & --- \\
    $\log q$ & $-3.238 \pm 0.125$ & $-2.930 \pm 0.130$ & $-2.558 \pm 0.047$ & --- \\
    $t_{0,2}$ (HJD$'$) & --- & --- & --- & $10064.63\pm  0.16$ \\
    $u_{0,2}$ & --- & --- & --- & $0.0116 \pm 0.0040$ \\
    $q_{f, I}$ & --- & --- & --- & $0.0232\pm 0.0037$ \\
    $I_{\rm S,KMTC}$ & $ 22.118\pm 0.274$ & $ 21.588\pm 0.212$ & $ 21.143\pm 0.050$ & $ 21.740\pm 0.179$ \\
    
    \hline\hline
  \end{tabular}
  \label{tab:parm-d}
\end{table*}

\begin{figure}
    \centering
    \includegraphics[width=0.47\textwidth]{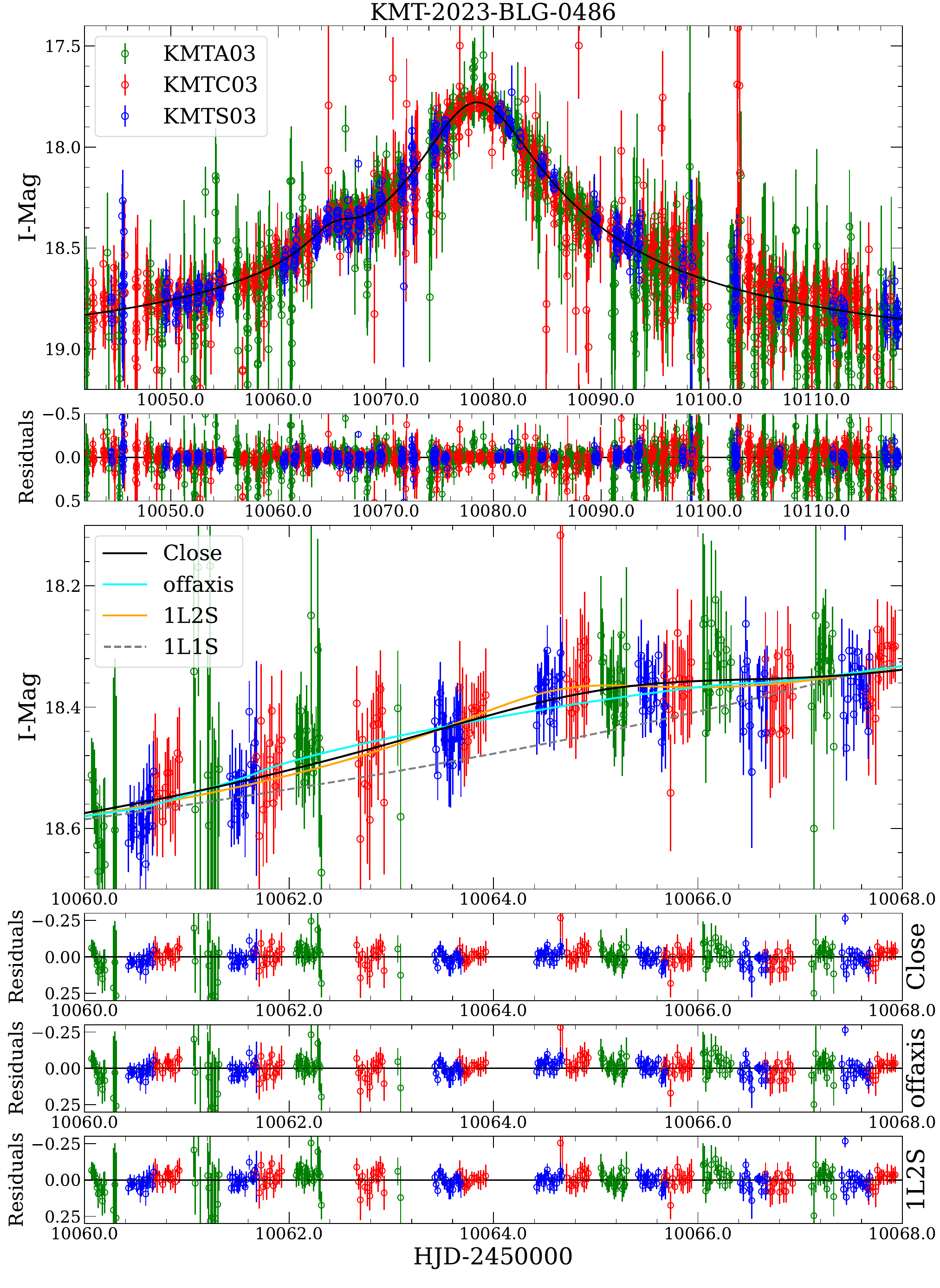}
    \caption{Light curves and models for \eventd. The symbols are similar to those in Figure \ref{fig:lc-a}.}
\label{fig:lc-d}
\end{figure}

Figure~\ref{fig:lc-d} shows the KMTNet light curve of \eventd\ obtained from its three observatory sites. The event largely follows a standard PSPL morphology, but exhibits an additional smooth, low-amplitude bump approximately 14 days prior to peak magnification. This anomaly persists for about five days with a peak amplitude of $\sim0.1$~mag. 

\begin{figure}
    \centering
    \includegraphics[width=0.47\textwidth]{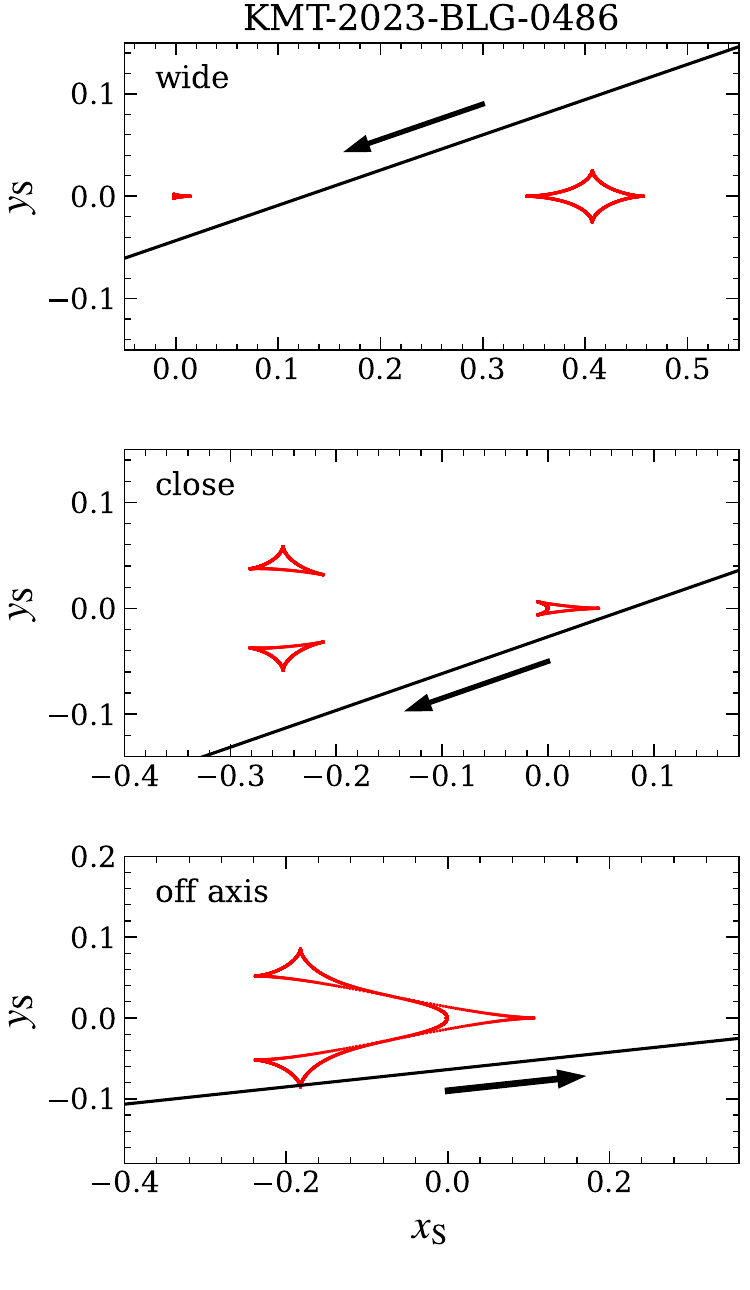}
    \caption{Geometries of the 2L1S models for \eventd. The symbols are similar to those in Figure \ref{fig:cau_abcef}.}
\label{fig:cau-d}
\end{figure}

Our grid search identifies three distinct local minima in parameter space, whose corresponding caustic geometries are illustrated in Figure~\ref{fig:cau-d}. In two of these solutions, which are labeled as ``Close'' and ``Wide'', the anomaly arises from a cusp approach associated with the central caustic. In the third solution, the bump is produced by an off-axis caustic crossing involving the central caustic, we therefore label this configuration as ``Off-axis''. Similar off-axis solutions have been reported in previous KMTNet events, such as KMT-2019-BLG-1216 \citep{2019_subprime}.

The posterior parameters derived from the MCMC analysis are summarized in Table~\ref{tab:parm-d}. Among the three solutions, the ``Close'' solution yields the best fit to the data, while the ``Wide'' solution is only marginally worse, with $\Delta\chi^2 = 0.3$. In contrast, the ``Off-axis'' solution is strongly disfavored, with $\Delta\chi^2 = 31.3$. This discrepancy is evident in the zoomed-in view of the anomaly shown in Figure~\ref{fig:lc-d}, in which the ``Off-axis'' model is worse to reproduce the observed bump, particularly near ${\rm HJD}^\prime \simeq 10065$ (where ${\rm HJD}^\prime \equiv {\rm HJD} - 2450000$). Moreover, the ``Off-axis'' solution requires significant finite-source effects, with a normalized source radius of $\rho = (2.85 \pm 0.36) \times 10^{-2}$. When combined with the extinction $A_I \sim 2.7$ based on the \citet{Nataf2013} extinction map and the source-property analysis routine described in Section~\ref{sec:lens}, this implies an exceptionally low lens-source relative proper motion of $\mu_{\rm rel} \sim 0.05~{\rm mas~yr}^{-1}$. Based on Equation~(9) of \citet{2018_subprime}, which is derived from the observed $\mu_{\rm rel}$ distribution of planetary microlensing events \citep{MASADA}, the probability of such a low proper motion is only $\sim7\times10^{-5}$. We therefore regard the ``Off-axis'' solution as physically implausible and exclude it from further consideration.

The 1L2S model is favored over the 2L1S model by $\Delta\chi^2 = 5.5$. Finite-source effects are not detected for either source, and therefore the 1L2S interpretation cannot be tested via the implied lens-source relative proper motion. In addition, the event lies in a region of relatively high extinction, with $A_I \sim 2.7$, and the anomaly is only weakly magnified, so the $V$-band data taken during the anomaly provide no useful constraint on the source color. Hence, the 1L2S interpretation cannot be tested using color information.

In conclusion, the lens-source system could be described by either a 2L1S or a 1L2S model. Because this is only a candidate planetary event, we do not pursue further analysis.

\subsection{\evente}

\begin{figure}
    \centering
    \includegraphics[width=0.47\textwidth]{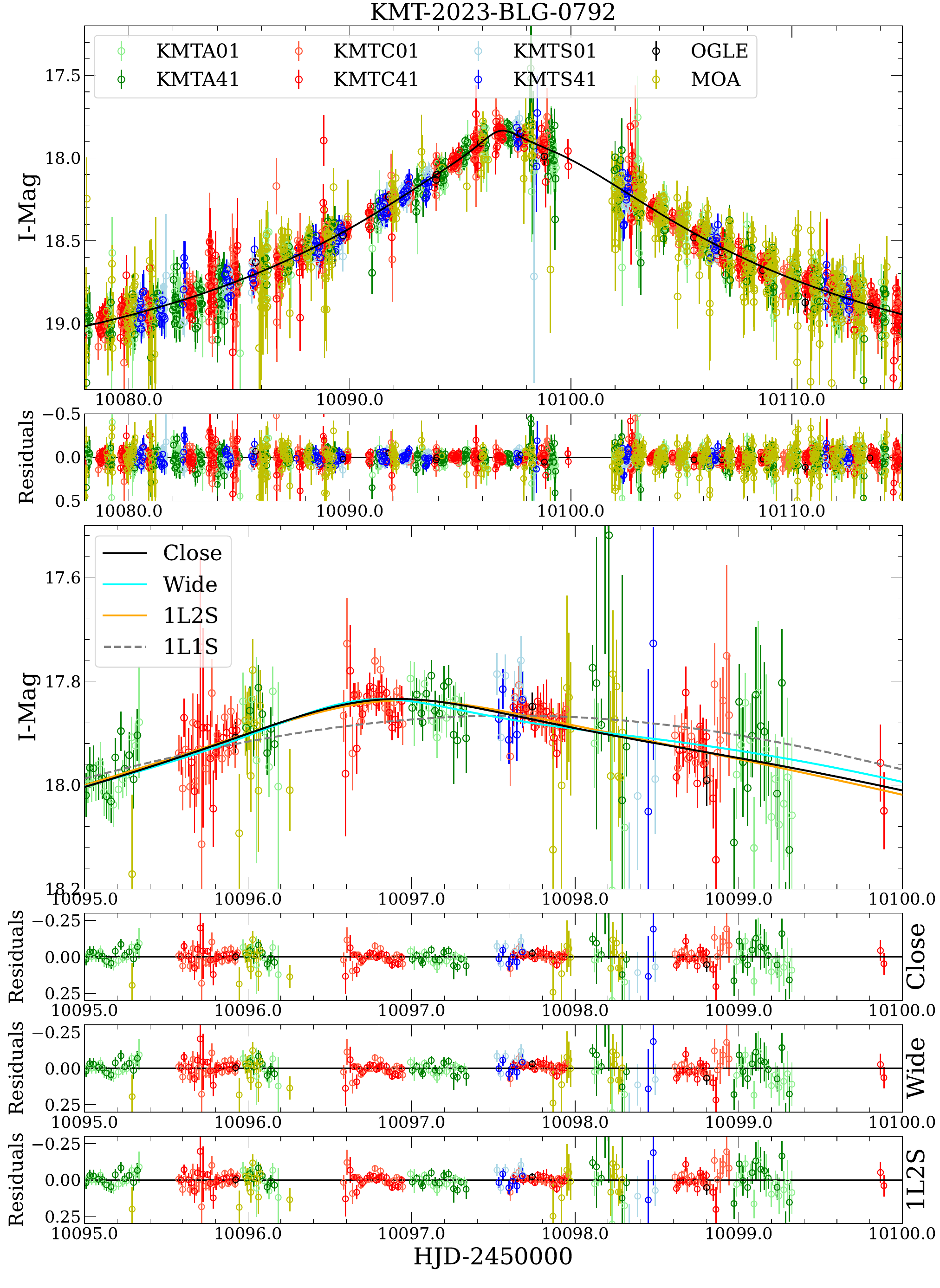}
    \caption{Light curves and models for \evente. The symbols are similar to those in Figure \ref{fig:lc-a}.}
\label{fig:lc-e}
\end{figure}

\begin{table*}[htb]
    \renewcommand\arraystretch{1.20}
    \caption{Lensing Parameters for \evente}
    \centering
        \begin{tabular}{c|ccc}
        \hline\hline
        Parameter & Wide & Close & 1L2S \\
        \hline
        $\chi^2/\mathrm{dof}$ 
        & 11445.8/11452 & 11430.7/11452 & 11426.1/11452 \\ \hline
        $t_0$ (HJD$^\prime$) 
        & $10097.561\pm0.021$ & $10097.569\pm0.018$ & $10097.643\pm0.025$ \\
        $u_0$ 
        & $0.0924\pm0.003$ & $0.0998\pm0.003$ & $ 0.116\pm0.006$ \\
        $t_{\rm E}$ (days) 
        & $54.05\pm1.53$ & $51.42\pm1.46$ & $49.32\pm1.54$ \\
        $\alpha$ (deg) 
        & $260.14\pm0.85$ & $261.03\pm0.71$ & ---\\
        $s$ 
        & $1.392\pm0.008$ & $0.710\pm0.008$ & --- \\
        $\log q$ 
        & $-2.99\pm0.04$ & $-2.60\pm0.05$ & --- \\
        $t_{0,2}$ (HJD$'$) & --- & --- & $10096.86\pm  0.05$ \\
        $u_{0,2}$ & --- & --- & $0.0229 \pm 0.0031$ \\
        $q_{f, I}$ & --- & --- & $0.0428\pm 0.0093$ \\
        $I_{\rm S,KMTC}$ 
        & $20.610\pm0.039$ & $20.520\pm0.068$ & $20.470\pm0.047$ \\
        \hline\hline
        \end{tabular}
        \label{tab:parm-e}
\end{table*}

Figure~\ref{fig:lc-e} shows the observed data for \evente, including six KMTNet data sets as well as data from the OGLE and MOA surveys. The anomaly appears as an asymmetric peak near the PSPL maximum and is covered by all data sets. The 2L1S grid search yields only a pair of solutions that exhibit the ``Close/Wide'' typology. As shown in Figure~\ref{fig:cau_abcef}, the anomaly is produced by a cusp approach to the central caustic. Table \ref{tab:parm-d} presents the lensing parameters. The ``Close'' solution provides the best fit, while the ``Wide'' solution is disfavored by $\Delta\chi^2 = 15.1$. According to our criterion, the ``Wide'' solution cannot be definitively ruled out.
Finite-source effects are not well constrained for either solution, with 1-$\sigma$ upper limits on $\rho$ exceeding 0.02. We therefore adopt the point-source model parameters.

The 1L2S model provides a slightly better fit than the 2L1S ``Close'' model, with $\Delta\chi^2 = 4.6$. The finite-source effect of the secondary source is consistent with a point-source model within $1\sigma$, so the 1L2S model cannot be tested using the ``kinematic argument''. In addition, due to the faintness of the event, the $V$-band data obtained over the anomaly have signal-to-noise ratios (SNR) below 10, which are insufficient to test the 1L2S interpretation through the measured color of the secondary source.

For this event, the parallax contours take the form of elongated ellipses. The constraint along the minor axis of the error ellipse is meaningful, with $\sigma(\pi_{\rm E,\parallel}) = 0.15$, where the parallel direction is approximately aligned with the Earth's acceleration. However, because the event can be modeled as either 2L1S or 1L2S, we do not pursue further analysis. 

\subsection{\eventf}

\begin{figure}

    \centering
    \includegraphics[width=0.47\textwidth]{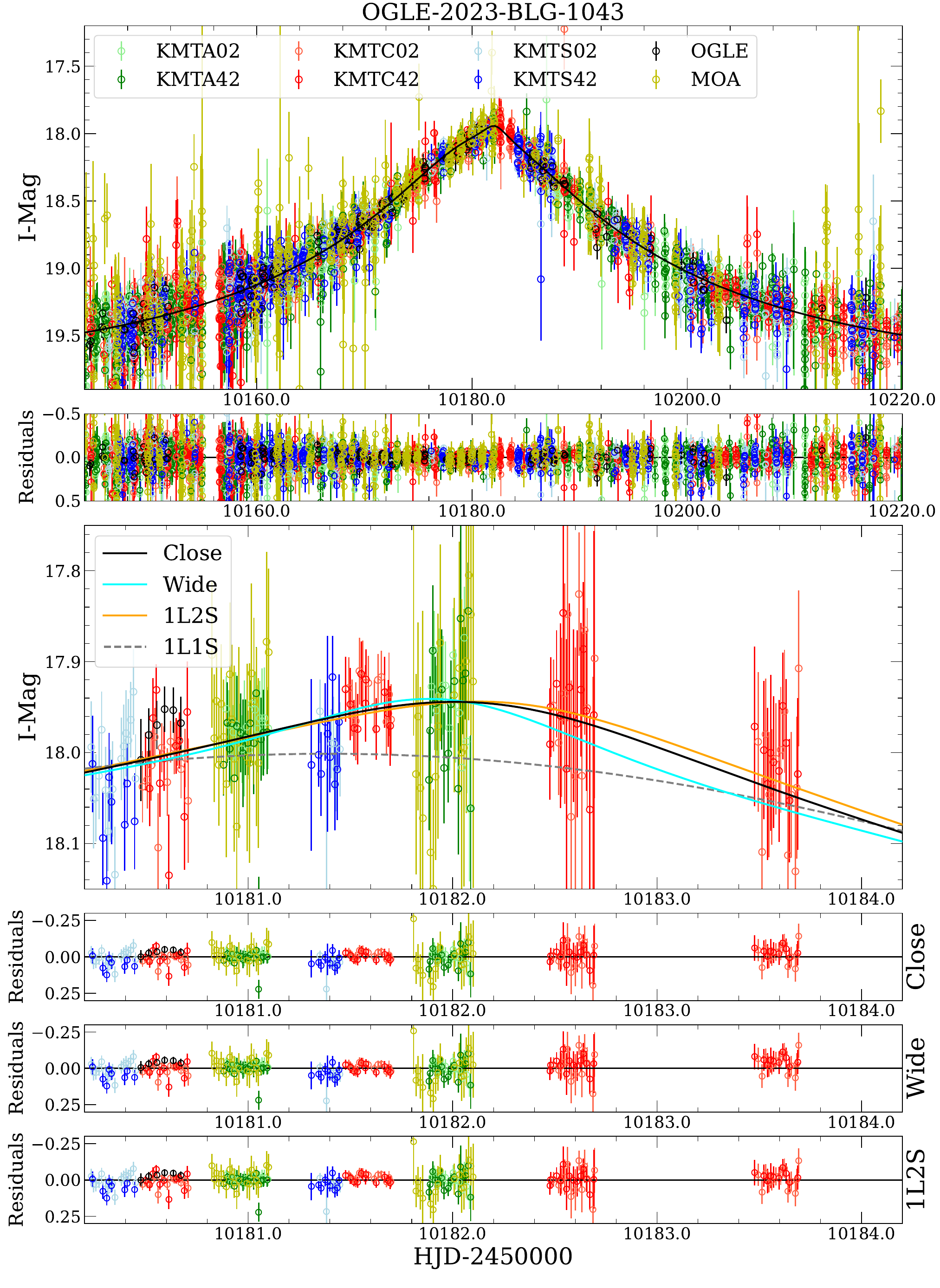}
    \caption{Light curves and models for \eventf. The symbols are similar to those in Figure \ref{fig:lc-a}.}
\label{fig:lc-f}
\end{figure}

\begin{table*}[htb]
    \renewcommand\arraystretch{1.20}
    \caption{Lensing Parameters for \eventf}
    \centering
        \begin{tabular}{c|ccc}
        \hline\hline
        Parameter & Wide & Close & 1L2S \\
        \hline
        $\chi^2/\mathrm{dof}$ 
        & 15672.0/15677 & 15657.0/15677 & 15649.0/15677 \\ \hline
        $t_0$ (HJD$^\prime$) 
        & $10181.37\pm0.02$ & $10181.36\pm0.02$ & $10181.27\pm0.04$ \\
        $u_0$ 
        & $0.150\pm0.04$ & $0.153\pm0.004$ & $ 0.172\pm0.007$ \\
        $t_{\rm E}$ (days) 
        & $44.42\pm0.84$ & $43.79\pm0.84$ & $42.38\pm0.98$ \\
        $\alpha$ (deg) 
        & $275.00\pm1.05$ & $276.71\pm1.02$ & ---\\
        $s$ 
        & $1.39\pm0.01$ & $0.76\pm0.02$ & --- \\
        $\log q$ 
        & $-2.896\pm0.053$ & $-2.583\pm0.05$ & --- \\
        $t_{0,2}$ (HJD$'$) & --- & --- & $10182.24\pm 0.10$ \\
        $u_{0,2}$ & --- & --- & $0.032 \pm 0.007$ \\
        $q_{f, I}$ & --- & --- & $0.032\pm 0.010$ \\
        $I_{\rm S,KMTC}$ 
        & $20.120\pm0.033$ & $20.093\pm0.031$ & $20.053\pm0.040$ \\
        \hline\hline
        \end{tabular}
        \label{tab:parm-f}
\end{table*}

The anomaly is also an asymmetric peak, as shown in Figure~\ref{fig:lc-f}. Compared to the 1L1S model (the grey dashed line), the anomaly appears as a bump-type feature occurring slightly after the 1L1S peak, and its overall morphology is similar to that of \evente. Therefore, both 2L1S and 1L2S models are considered. For the 2L1S interpretation, a pair of ``Close/Wide'' solutions is found. The ``Close'' solution provides the best fit, while the ``Wide'' solution is disfavored by $\Delta\chi^2 = 15.0$, as listed in Table~\ref{tab:parm-f}. The anomaly is also produced by a cusp approach to the central caustic, as shown in Figure~\ref{fig:cau_abcef}.

The 1L2S model is favored by $\Delta\chi^2 = 8.0$ relative to the 2L1S ``Close'' model. For the same reasons as in \evente, the 1L2S model cannot be rejected based on the derived lens-source relative proper motion or the color of the putative secondary source. Therefore, for this event, both 2L1S and 1L2S models can explain the observations, and we do not pursue further analysis.

\section{Source and Lens Properties}\label{sec:lens}

\begin{figure*}[htb] 
    \centering
    \includegraphics[width=\textwidth]{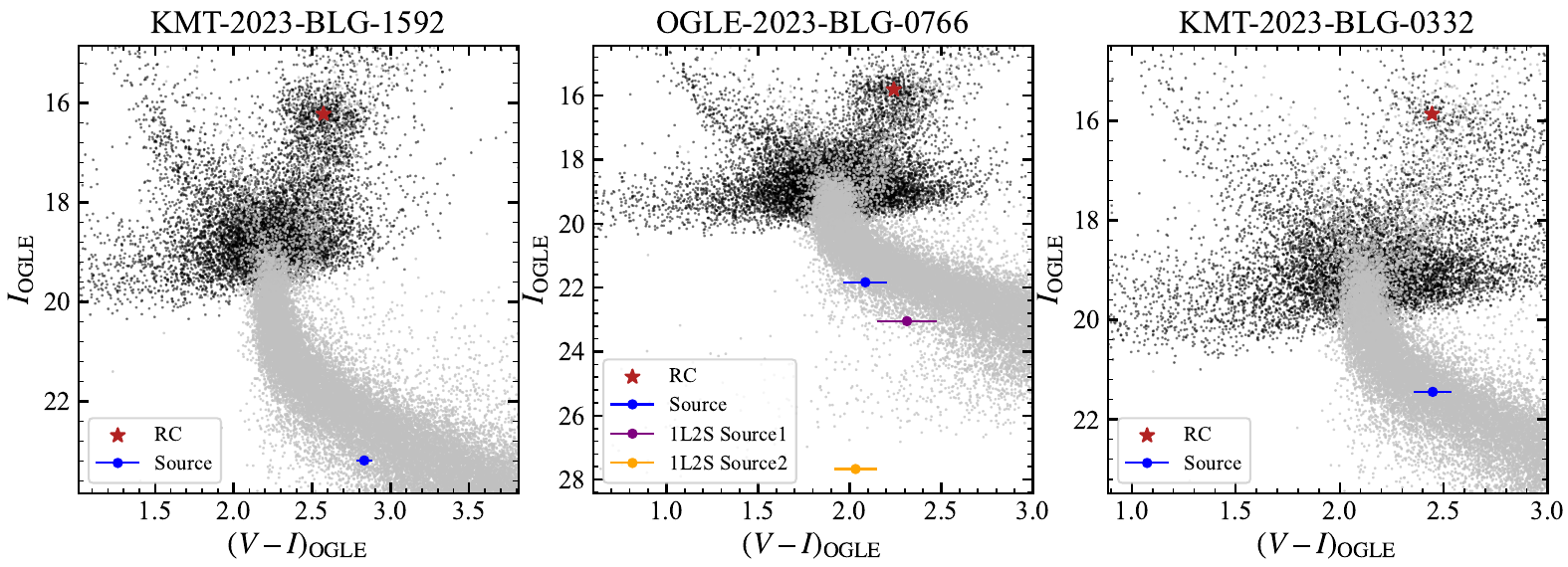}
    \caption{Color-magnitude diagrams of the three unambiguous 2L1S events, constructed using OGLE-III stars within a $2'$ radius centered on each event \citep{OGLEIII}. In each panel, the red asterisk marks the centroid of the red clump, and the blue dot indicates the 2L1S source star. For \eventb, the two 1L2S source stars are also shown. The gey dots represent the {\it HST} CMD from \cite{HSTCMD}. The centroid of the red-giant clump in the {\it HST} CMD, $(V - I, I)_{\rm cl, HST} = (1.62, 15.15)$ \citep{MB07192}, is aligned with that of the OGLE-III CMD.}
    \label{fig:cmd}
\end{figure*}

\begin{table*}[htb]
\centering
\small
\setlength{\tabcolsep}{6pt}
\renewcommand{\arraystretch}{1.2}
\caption{CMD parameters, source properties and derived $\theta_{\rm E}$ and $\mu_{\rm rel}$}
\begin{tabular}{l | cc | cc | c}
\hline\hline

& \multicolumn{2}{c|}{\textbf{\eventa}}&\multicolumn{2}{c|}{\textbf{\eventb}}
& \textbf{\eventc} \\
Parameter
& Wide & Close
& Wide & Close
&  \\

\hline
$I_{\rm RC}$
& 16.22 $\pm$ 0.02
& $\xleftarrow{}$
& 15.81$\pm $ 0.03
& $\xleftarrow{}$
& 15.86 $\pm$ 0.06\\

$(V-I)_{\rm RC}$
& 2.57 $\pm$ 0.01
& $\xleftarrow{}$
& 2.24$\pm$ 0.01
& $\xleftarrow{}$
& 2.44 $\pm$ 0.01\\

$I_{\rm RC,0}$
& 14.40 $\pm$ 0.04
& $\xleftarrow{}$
& 14.37$\pm$ 0.04
& $\xleftarrow{}$
& 14.34 $\pm$ 0.04\\

$(V-I)_{\rm RC,0}$
& 1.06 $\pm$ 0.03
& $\xleftarrow{}$
& $\xleftarrow{}$
& $\xleftarrow{}$
& $\xleftarrow{}$\\

$I_{\rm S}$
& 23.16 $\pm$ 0.27
& 23.18 $\pm$ 0.25
& 21.92 $\pm$ 0.20
& 21.83 $\pm$ 0.13
& 21.44 $\pm$ 0.20 \\

$(V-I)_{\rm S}$
& 2.83 $\pm$ 0.05
& $\xleftarrow{}$
& 2.08$\pm$ 0.12
& $\xleftarrow{}$
& 2.45 $\pm$ 0.11\\

$I_{\rm S,0}$
& 21.33 $\pm$ 0.27
& 21.36 $\pm$ 0.25
& 20.48 $\pm$ 0.20
& 20.40 $\pm$ 0.14
& 19.92 $\pm$ 0.21 \\

$(V-I)_{\rm S,0}$
& 1.32 $\pm$ 0.06
& $\xleftarrow{}$
& 0.90$\pm$ 0.12
& $\xleftarrow{}$
& 1.07 $\pm$ 0.12\\

$\theta_\ast$ ($\mu$as)
& 0.298 $\pm$ 0.043
& 0.294 $\pm$ 0.040
& 0.307 $\pm$ 0.045
& 0.319 $\pm$ 0.041
& 0.457 $\pm$ 0.068 \\

$\theta_{\rm E}$ (mas)
& 0.425 $\pm$ 0.112
& 0.478 $\pm$ 0.127
& 0.318 $\pm$ 0.069
& 0.319 $\pm$ 0.042
& --- \\

$\mu_{\rm rel}$ (mas\,yr$^{-1}$)
& 1.930 $\pm$ 0.654
& 1.901 $\pm$ 0.667
& 1.889 $\pm$ 0.453
& 2.113 $\pm$ 0.479
& --- \\

\hline\hline
\end{tabular}
\label{tab:cmd}
\end{table*}

\subsection{Preamble}

We determine the lens physical properties of three unambiguous 2L1S events, \eventa, \eventb, and \eventc, in this section. Following the procedure outlined in Section~\ref{model_preamble}, we first describe the general methodology. 

The lens mass $M_{\rm L}$ and distance $D_{\rm L}$ are expressed in terms of the angular Einstein radius and the microlensing parallax \citep{Gould1992, Gould2000} as
\begin{equation}\label{eq:mass}
M_{\rm L} = \frac{\thetae}{\kappa \pie}; \qquad
D_{\rm L} = \frac{\mathrm{au}}{\pie \thetae + \pi_{\rm S}},
\end{equation}
where $\pi_{\rm S}$ denotes the source parallax. The angular Einstein radius can be obtained from $\thetae = \theta_*/\rho$. To evaluate $\theta_*$, we place the source on the $V-I$ versus $I$ color-magnitude diagram (CMD; \citealt{Yoo2004}) constructed from OGLE-III stars within a $2'$ radius centered on the event \citep{OGLEIII}. The centroid of the red clump (RC), ${(V - I, I)}_{\rm RC}$, is derived using the method of \citet{Nataf2013}, and its intrinsic color and magnitude, $(V - I, I)_{\rm RC,0}$, are adopted from \cite{Bensby2013} and Table~1 of \cite{Nataf2013}. Then, the total extinction in the $I$ band and the reddening toward each event are derived as
\begin{equation}
  [E(V-I), A_I] =  [(V - I, I)_{\rm RC}] - [(V - I, I)_{\rm RC,0}].
\end{equation}

For each event, the apparent source magnitude is obtained from the light-curve modeling and calibrated to the OGLE-III photometric system using common field stars between KMTNet and OGLE-III. For \eventa\ and \eventb, the source color $(V-I)_{\rm S}$ is determined from a regression of the KMTNet $V$-band and $I$-band fluxes as a function of magnification and then calibrated to the OGLE-III color scale. The intrinsic source magnitude and color are thus derived as
\begin{equation}
    [(V-I), I]_{\rm S,0} = [(V-I), I]_{\rm S} - [E(V-I), A_I].
\end{equation}
For \eventc, because of the low SNR in the $V$ band, we determine the source color by aligning the {\it Hubble} Space Telescope ({\it HST}) CMD from \cite{HSTCMD} to the OGLE-III CMD using the measured $I_{\rm cl}$ of the red clump. The source color is then estimated from {\it HST} field stars within 5-$\sigma$ of the source brightness. Applying the color/surface-brightness relation of \citet{Adams2018}, we obtain $\theta_*$, which leads to measurements of $\theta_{\rm E}$ and $\mu_{\rm rel}$. 

Figure~\ref{fig:cmd} displays the CMDs of the three events, and Table~\ref{tab:cmd} summarizes the corresponding CMD parameters, source properties, $\theta_{\rm E}$, and $\mu_{\rm rel}$ for all solutions.

Among the three events, none of events have useful constraints on the microlensing parallax. We therefore infer the physical parameters of the systems through a Bayesian analysis that adopts a Galactic model as the prior. The model assumptions and computational framework follow \cite{Yang2021_GalacticModel}, including the assumption that the planet occurrence rate does not depend on host-star properties such as mass. Additional details are given in that work. 

\begin{figure*}[htb] 
    \centering
    \includegraphics[width=0.9\textwidth]{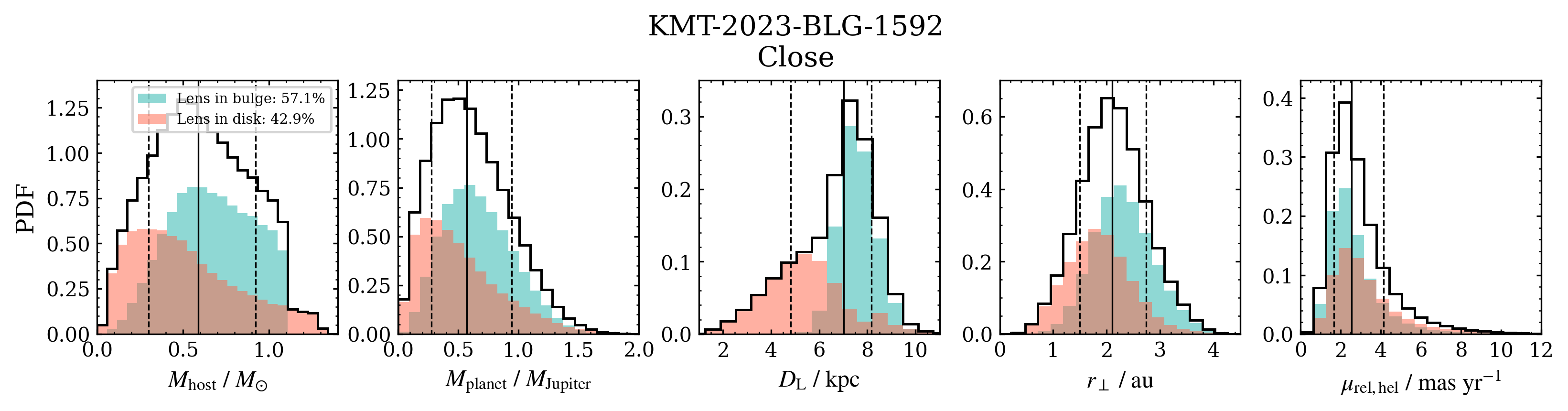}
    \includegraphics[width=0.9\textwidth]{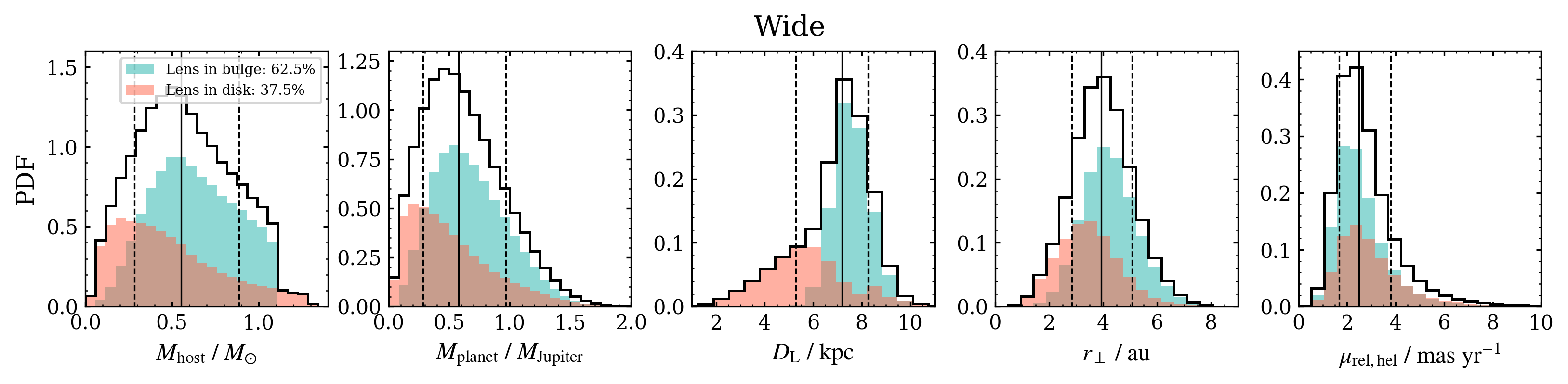}
    \includegraphics[width=0.9\textwidth]{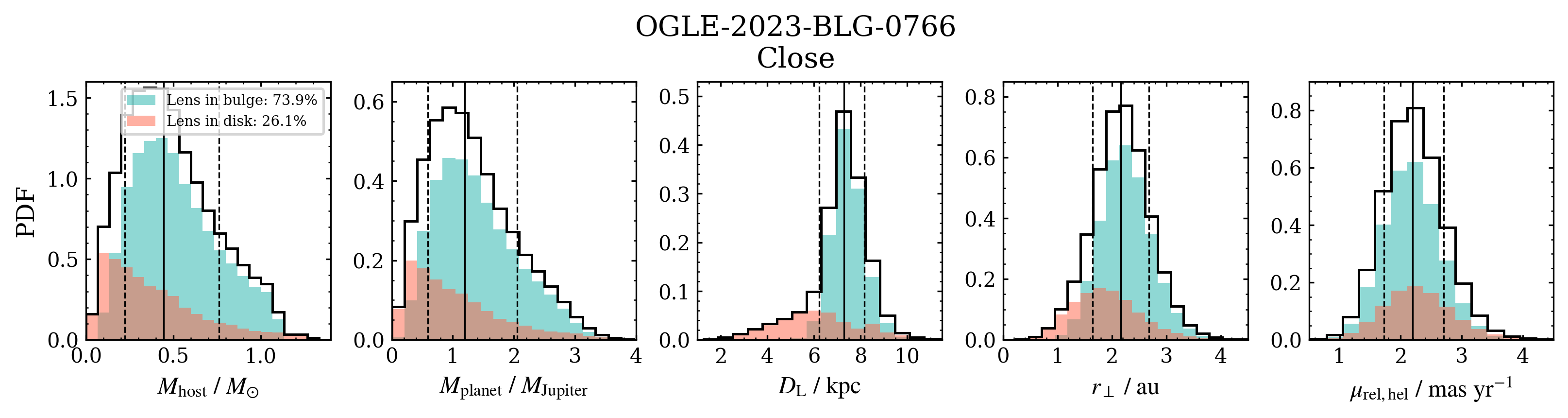}
    \includegraphics[width=0.9\textwidth]{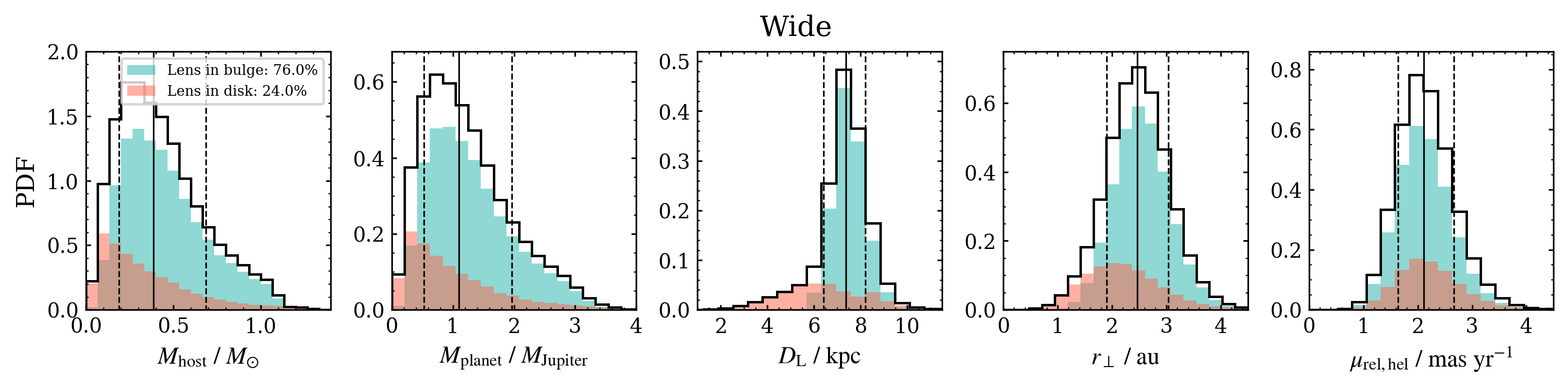}
    \includegraphics[width=0.9\textwidth]{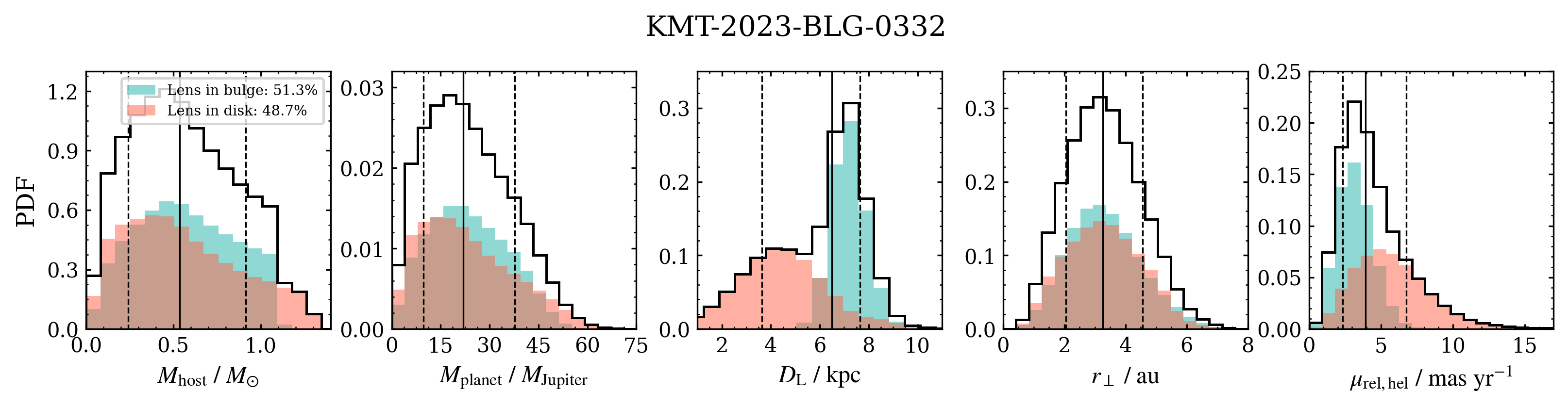}
    \caption{Posterior distributions from the Bayesian analysis for the host mass $M_{\rm host}$, planetary mass $M_{\rm planet}$, lens distance $D_{\rm L}$, projected planet-host separation $r_\bot$, and the heliocentric lens-source relative proper motion $\mu_{\rm rel,hel}$. In each panel, the solid black curve denotes the median value, while the dashed black curves represent the 15.9\% and 84.1\% credible intervals. The contributions from disk and bulge lens populations are illustrated in green and red, respectively.}
    \label{fig:baye}
\end{figure*}

The posterior distributions from the Bayesian analysis are summarized in Table~\ref{tab:baye} and illustrated in Figure~\ref{fig:baye}. These include the host mass $M_{\rm host}$, planet mass $M_{\rm planet}$, lens distance $D_{\rm L}$, projected lens separation $r_{\perp}$, and the heliocentric lens-source relative proper motion $\mu_{\rm hel, rel}$.

\begin{table*}[htb]
\renewcommand\arraystretch{1.25}
\centering
\caption{Lensing Physical Parameters from Bayesian Analyses}
\begin{tabular}{c c | c c c c c}
\hline
\hline
Event & Model 
& $M_\mathrm{host}\,(M_\odot)$ 
& $M_\mathrm{planet}\,(M_{\rm jupiter})$ 
& $D_\mathrm{L}\,(\mathrm{kpc})$ 
& $r_\bot\,(\mathrm{au})$ 
& $\mu_\mathrm{rel,hel}\,(\mathrm{mas\,yr^{-1}})$ \\
\hline

\multirow{2}{*}{KMT-2023-BLG-1592}
& Close
& $0.59_{-0.29}^{+0.33}$
& $0.57_{-0.29}^{+0.37}$
& $7.0_{-2.2}^{+1.1}$
& $2.1_{-0.6}^{+0.6}$
& $2.54_{-0.88}^{+1.59}$ \\
& Wide
& $0.55_{-0.27}^{+0.33}$
& $0.58_{-0.29}^{+0.39}$
& $7.2_{-1.9}^{+1.1}$
& $3.9_{-1.1}^{+1.1}$
& $2.48_{-0.80}^{+1.31}$ \\

\hline

\multirow{2}{*}{OGLE-2023-BLG-0766}
& Close
& $0.45_{-0.22}^{+0.32}$
& $1.20_{-0.60}^{+0.86}$
& $7.3_{-1.1}^{+0.9}$
& $2.2_{-0.5}^{+0.5}$
& $2.19_{-0.46}^{+0.51}$ \\
& Wide
& $0.39_{-0.20}^{+0.30}$
& $1.10_{-0.57}^{+0.87}$
& $7.4_{-0.9}^{+0.9}$
& $2.5_{-0.6}^{+0.6}$
& $2.11_{-0.47}^{+0.55}$ \\

\hline

KMT-2023-BLG-0332
& ---
& $0.54_{-0.30}^{+0.38}$
& $22.0_{-12.2}^{+15.7}$
& $6.5_{-2.8}^{+1.2}$
& $3.2_{-1.2}^{+1.3}$
& $3.93_{-1.60}^{+2.82}$ \\

\hline
\hline
\end{tabular}
\label{tab:baye}
\end{table*}

\subsection{\eventa}

The intrinsic source color and magnitude are $[(V-I), I]_{\rm S,0} = [1.10 \pm 0.04, 20.68 \pm 0.17]$. Based on its position relative to the {\it HST} field stars in Figure~\ref{fig:cmd}, the source is likely a metal-poor K dwarf located in the bulge. The lens-source relative proper motion, $\sim 1.9$~mas~yr$^{-1}$, is normal for a planetary event \citep{MASADA}.

The Bayesian analysis indicates that the host is likely an M- or K-dwarf. The planetary mass is $\sim 0.6~M_{\rm Jupiter}$, and the projected planet-host separation is $\sim 4$ au for the ``Wide'' solution and $\sim 2$ au for the ``Close'' solution. Adopting a water snow-line scaling of $a_{\rm SL} = 2.7 (M/M_\odot)$~au \citep{snowline}, the planet lies well beyond the snow line in the ``Wide'' solution and may reside near the snow line in the ``Close'' solution.

\subsection{\eventb}\label{sec:phy_eventb}

For this event, the CMD also helps exclude the 1L2S model. As shown in Figure~\ref{fig:cmd}, we plot the two 1L2S source stars. The hypothetical secondary source would be extremely faint, $I \sim 27.6$, and significantly bluer than the main-sequence branch. Such a position in the CMD is physically implausible, rendering the 1L2S interpretation unlikely.

The Bayesian analysis indicates that the host star is most likely an M-dwarf. The planet has approximately Jupiter mass, and the projected planet-host separation is $\sim 2.3$ au. Therefore, this gas-giant planet is also beyond the snow line.

\subsection{\eventc}

For this event, $\rho$ is not usefully constrained, and therefore no meaningful constraints can be placed on $\theta_{\rm E}$ or $\mu_{\rm rel}$. Nevertheless, we derive the source properties from the CMD, as this information may assist in planning high-resolution follow-up observations. The source is $5.6$ mag fainter than the red clump, suggesting that it is likely a K-dwarf and therefore unlikely to be a bright blend relative to the lens in future high-resolution imaging.

The Bayesian analysis indicates that the host is also likely an M- or K-dwarf. The companion mass is $22.0^{+15.7}_{-12.2}~M_{\rm Jupiter}$, implying that it is most likely a low-mass brown dwarf. The lens-source relative proper motion is $3.93^{+2.82}_{-1.60}$~mas~yr$^{-1}$.

\section{Discussion}\label{sec:dis}

We have presented the observations and analysis of six anomalous microlensing events located in the KMTNet prime field. Among them, three are securely classified as 2L1S events, including two planetary systems and one brown dwarf system, while the remaining three exhibit the 2L1S/1L2S degeneracy. For the three confirmed 2L1S events, the anomaly in \eventa\ was initially identified through by-eye searches, whereas the anomalies in all three 2L1S/1L2S cases were detected by the AnomalyFinder system.

Since the 1L2S degeneracy in planetary microlensing events was first identified by \cite{Gaudi1998}, it has remained the most common ambiguity in planetary interpretations (e.g., \citealt{MB12486,KB171119}). In constructing the largest microlensing planetary statistical sample \citep{OB160007}, 11 of 63 candidate planets were excluded due to the 2L1S/1L2S degeneracy\footnote{\cite{OB160007} reported nine such cases, and two additional events, OGLE-2018-BLG-1554 \citep{2018_prime} and OGLE-2019-BLG-0344 \citep{2019_subprime}, excluded for degenerate binary-star solutions, also exhibit 1L2S degeneracy.}, corresponding to a degeneracy fraction of 17.5\%. Of these 11 events, eight were first identified by AnomalyFinder and three via by-eye searches, yielding an AnomalyFinder discovery fraction of 72.7\%, substantially higher than the $\sim 33\%$ rate for unambiguous planets \citep{2017_subprime}. This trend is expected because AnomalyFinder is more sensitive to subtle anomalies, for which the lower SNRs makes them harder to distinguish from 1L2S solutions using light curves' $\chi^2$, color, or kinematic constraints. 

With the upcoming Nancy Grace Roman Space Telescope \citep{MatthewWFIRSTI} and the Earth 2.0 Microlensing Telescope (ET) \citep{CMST,ET}, the number of detected planetary microlensing events per year is expected to increase dramatically. The systematic discovery of planets by AnomalyFinder underscores its importance for future survey data analysis. At the same time, the high 2L1S/1L2S degeneracy rate within the AnomalyFinder sample calls for particular caution when interpreting planetary signals with bump-type features.

Our analysis also identifies a brown dwarf event with $\log q = -1.4$. At present, only one homogeneous microlensing sample probes the brown dwarf mass-ratio regime, $-1.5 < \log q < -1.0$, based on Wise, OGLE, and MOA data \citep{Wise}, although the number of events in that sample remains small. Current KMTNet statistical studies primarily focus on planetary companions with $\log q < -1.5$. Ongoing analyses of brown-dwarf systems discovered through by-eye searches are reported in \citep{Han_3BDs,HanBD1618,HanBD1820,HanBD21,HanBD2223}. A future systematic study of brown dwarf events identified by AnomalyFinder could extend the KMTNet mass-ratio function toward $\log q \sim -1$.

\bigskip

Z.L., H.L., W.Z., H.Y., Y.T., J.Z. and S.M. acknowledge support by the National Natural Science Foundation of China (Grant No. 12133005). This research has made use of the KMTNet system operated by the Korea Astronomy and Space Science Institute (KASI) at three host sites of CTIO in Chile, SAAO in South Africa, and SSO in Australia. Data transfer from the host site to KASI was supported by the Korea Research Environment Open NETwork (KREONET). This research was supported by KASI under the R\&D program (Project No. 2025-1-830-05) supervised by the Ministry of Science and ICT. The OGLE project has received funding from the Polish National Science Centre grant OPUS 2024/55/B/ST9/00447 awarded to A.U. The MOA project is supported by JSPS KAKENHI Grant Number JP24253004, JP26247023,JP16H06287 and JP22H00153. This work is part of the ET space mission which is funded by the China's Space Origins Exploration Program. H.Y. acknowledges support by the China Postdoctoral Science Foundation (No. 2024M762938). Y.S. acknowledges support from BSF Grant No. 2020740. R.P. acknowledges support by the Polish National Agency for Academic Exchange grant ``Polish Returns 2019.''

\software{pySIS \citep{pysis,Yang_TLC,Yang_TLC2}, numpy \citep{numpy}, emcee \citep{emcee2,emcee}, Matplotlib \citep{Matplotlib}, SciPy \citep{scipy}}, VBBL \cite{Bozza2010,Bozza2018,VBMicrolensing2025}

\bibliography{Zang.bib}

\end{CJK*}
\end{document}